\documentclass[11pt,a4paper]{article}
\usepackage{jheppub}
\usepackage{comment}

\def\be{\begin{equation}}
\def\ee{\end{equation}}
\def\ba{\begin{eqnarray}}
\def\ea{\end{eqnarray}}
\def\bi{\begin{itemize}}
\def\ei{\end{itemize}}

\def\sint{\textstyle{\int}}
\def\xh{\hat{x}}
\def\qh{\hat{q}}
\def\ph{\hat{p}}
\def\wb{\bar{w}}
\def\zb{\bar{z}}
\def\w{\omega}
\def\t{\tau}
\def\dpp{\widetilde{d p}\,}

\def\D{\mathcal{D}}
\def\H{\mathcal{H}}

\def\e{\epsilon}
\def\M{\mathcal{M}}
\def\Qs{Q^{\text{soft}}}
\def\Qh{Q^{\text{hard}}}
\def\Kwb{\mathbb{K}_{\wb}}
\def\T{\mathbb{T}}

\title{Asymptotic charges in massless QED revisited:\\ A view from Spatial Infinity}

\author[a]{Miguel Campiglia} 
\author[b]{Alok Laddha} 
\affiliation[a]{Instituto de F\'isica, Facultad de Ciencias,  Montevideo 11400, Uruguay}
\affiliation[b]{Chennai Mathematical Institute, Siruseri, Chennai, India}

\emailAdd{campi@fisica.edu.uy}
\emailAdd{aladdha@cmi.ac.in}

\abstract{
Hamada and Shiu have recently shown that tree level amplitudes in QED satisfy an infinite hierarchy of soft photon theorems, the first two of which are Weinberg and Low's theorems respectively. In this paper we propose that in tree level  massless QED, this entire hierarchy is equivalent to a hierarchy of (asymptotic) conservation laws.  We prove the equivalence explicitly for the case of sub-subleading soft photon theorem and give substantial evidence that the equivalence continues to hold for the entire hierarchy. Our work also brings out the (complimentary) relationship between the asymptotic charges associated to soft theorems and the well known Newman-Penrose charges.
}

\begin{document}

\maketitle

\section{Introduction and outline}

The relationship between soft theorems in gauge theories and gravity and asymptotic symmetries in these theories is an active area of investigation. 
The essential idea is simple and can be understood without referring to specific theories. Given a large gauge transformation\footnote{In this paper we will use the expressions ``asymptotic symmetry'' and ``large gauge transformation'' interchangeably.} which is parametrized by a gauge parameter $\epsilon$ on a cross section $S^{2}$ of null infinity ${\cal I}$,  %(in certain cases, $\epsilon$ turns out to be a vector field on the $S^{2}$), 
the soft theorems are statements regarding conservation laws:
\begin{equation}
Q_{+}[\epsilon^{+}]\ =\ Q_{-}[\epsilon^{-}]
\end{equation}
where $\epsilon^{\pm}$ parametrizes large gauge transformations at ${\cal I}^{\pm}$ and are  identified via anti-podal matching conditions. $Q_{\pm}[\epsilon^{\pm}]$ are charges which are evaluated on the celestial sphere $S^{2}$ at $u\ =\ -\infty\ ({\cal I}^{+}_{-})$ and $v\ =\ +\infty\ ({\cal I}^{-}_{+})$ respectively. 

Thus, a rather natural question to ask is, if there indeed are such infinity of conservation laws in theories like Electrodynamics, should we not be able to derive them in classical theory? 
First step in this direction was taken in \cite{eyhe} where it was shown that the infinity of charges associated to large $U(1)$ gauge transformations  were indeed conserved in classical theory (from previous works \cite{strom0,stromprahar} these conservation laws in quantum theory were known to be equivalent to soft photon theorem). This conservation law was derived by analyzing the equations of the theory at spatial infinity. By  considering a compactification scheme where spatial infinity is a hyperboloid ${\cal H}$ (a three dimensional Lorentzian de Sitter space in fact)  as opposed to a point, one could  relate fields at $u\ =\ -\infty$ and $v\ =\ +\infty$ by using the field equations. 

We revisit this idea below and show that it leads to  (classical) conservation law for charges associated to leading as well as subleading soft photon theorem.\footnote{Subleading soft photon theorem is non-universal \cite{elvang}, however this most general form of subleading theorem can be understood in terms of asymptotic symmetries \cite{prahar}.} However this derivation leads  to an interesting consequence. That, if we restrict ourselves to a suitable subset in the set of all radiative data, there is in fact an entire tower of conservation laws that can be shown to be valid.  A priori, it is not clear what the Ward identities corresponding to these hierarchy of conservation laws imply in quantum theory.

We thus have the following question to answer:

{\bf (A)} If such infinite tower of charges are conserved, we would expect (along the lines of leading and sub-leading soft photon theorems) an infinite hierarchy of soft theorems. Do such theorems exist? 

As it turns out, from the side of soft theorems, a complimentary puzzle already existed:

{\bf (B)} In \cite{shiu,llz}, it was shown that in fact for tree level scattering amplitudes in QED, there do exist an infinity of soft theorems.  From the perspective of these soft theorems, a natural question to ask would be, just as leading and sub-leading soft theorems are Ward identities for certain asymptotic symmetries. Is the same true for the higher order theorems?  

In this paper we present substantial evidence that questions (A) and (B) mutually answer each other. That is, the conservation of asymptotic charges (beyond the ones associated to sub-leading soft theorem) are equivalent to sub-$n$ ($n > 1$) soft theorems.

The outline of the paper is as follows. In section \ref{two}, we revisit the asymptotic analysis of Maxwell's equations at null infinity. For simplicity, we consider charged massless scalar fields coupled fo $U(1)$ gauge fields but our analysis can be generalized to the situation when charged fields are massive and are scalars or Fermions. We then revisit the derivation of asymptotic charges associated to  leading and sub-leading soft photon theorems.  Our derivation of these charges is along the lines of  \cite{strom1,condemao} in  that  we first obtain them as integrals over $S^{2}$ at $u\ =\ -\infty$, and then show that they can be written as fluxes over ${\cal I}$. The reason for revisiting these charges is that our derivation is amenable to a direct generalization to an infinite hierarchy of further conservation laws. %In section \ref{four} we write these new family of charges as fluxes over null infinity. 
In section \ref{three} we review the infinity of soft theorems derived in \cite{shiu,llz}  for tree level scattering amplitudes and show how these theorems can be written in terms of Ward identities. In section \ref{conj-equi} we argue that these Ward identities precisely correspond to the hierarchy of conservation laws proposed above. Finally in section \ref{class-cons}, we show how this infinite hierarchy of conservation laws are indeed true in classical theory, provided one restricts attention to certain subset of radiative data that is compatible with tree-level scattering.  We end with certain speculations and future directions.

\section{Maxwell equations at ${\cal I}^{+}$ and asymptotic charges} \label{two}

In this section, we review the asymptotic expansion of Maxwell fields at null infinity and study Maxwell's equations in an $\frac{1}{r}$ expansion.   We will work in terms of self-dual fields since this simplifies the  field equations (bringing them into a form equivalent to the Newman-Penrose formulation \cite{npcharges}) and because  the charges associated to (positive) negative helicity soft photon theorem can be written in terms of  (anti) self-dual fields. Our first aim will be to understand how the self-dual  field can be determined order by order in $\frac{1}{r}$ in terms of free data at ${\cal I}^{+}$. At each order there will appear new `integration constants' that will be interpreted as  asymptotic charges under the assumption of certain strong $|u| \to \infty$ fall-offs. We will discuss in detail the first two set of asymptotic charges and show how they correspond to the charges associated to the leading and subleading  (negative helicity) soft photon theorems. We will finally discuss the relation with Newman-Penrose charges.

We consider $U(1)$ gauge field minimally coupled to massless scalar field. In order to analyze the behavior of the radiation fields at null infinity, we rewrite the equations of motion in terms of retarded coordinates $ (u, r, z, \zb)$ as, 

\ba
r^2 j_r & = & - \partial_r (r^2 F_{ru}) +D^{A}F_{rA} \label{eqr}\\
r^2 j_u & = & - \partial_r (r^2 F_{ru})+r^2 \partial_u F_{ru}+ D^A F_{uA} \label{equ}\\
j_A &=& \partial_r (F_{uA}-F_{rA}) + \partial_u F_{rA} + r^{-2}D^{B}F_{AB}. \label{eqA}
\ea
where sphere indices are raised with the unit-sphere metric $\gamma_{AB}$. These should be supplemented with Bianchi identities $\partial_{[a} F_{bc]}=0$.

An alternative description can be given in terms of the self-dual field strength as follows. Define, dual, self-dual and anti-self-dual fields by:
\be
\tilde{F}_{ab} := \frac{1}{2} \e_{abcd}F^{cd}, \quad F^\pm_{ab}:= F_{ab} \mp i \tilde{F}_{ab}.
\ee
Then the self-dual field satisfies the equations
\be
j_b= \nabla^b F^+_{ab} , \quad  \tilde{F}^+_{ab} =  i F^{+}_{ab} . \label{Fpluseqs}
\ee

From the perspective of soft theorems, the self-dual and anti-self dual equations are better suited than the ordinary Maxwell's equations in terms of real fields since  the positive and negative helicity soft photon theorems are related to charges constructed from the $F^\mp_{ru}$ fields. For definitiveness we work with self-dual field $F^+_{ab}$ whose quantization gives single particle states associated to negative helicity photons. 

The standard fall-off conditions (in $\frac{1}{r}$) which accommodates radiative data are\footnote{To simplify notation we will omit the $+$ superscript in the coefficients for $F^+_{ru}$.}

\ba
F^+_{ru}(r,u,\xh) & = & \frac{1}{r^2} \sum_{n=0} \frac{1}{r^n}\overset{n}{F}_{ru}(u,\xh) \label{rexpFru}\\
j_z(r,u,\xh) & = & \frac{1}{r^2} \sum_{n=0} \frac{1}{r^n}\overset{n}{j}_z(u,\xh) \label{rexpjz} \\
j_{r}(r,u,\xh) & = & \frac{1}{r^4} \sum_{n=0} \frac{1}{r^n}\overset{n}{j}_{r}(u,\xh)  \label{rexpjr} 
\ea

As shown in appendix \ref{maxwell-selfdual}, Maxwell equations for self-dual fields gives rise to the following recursive equation for $\overset{n}{F}_{ru}$:

\be
2(n+1) \partial_u \overset{n+1}{F}_{ru} + (\Delta + n(n+1)) \overset{n}{F}_{ru} = -2 D^z \overset{n}{j}_z + 2 \partial_u  \overset{n}{j}_r + (n+1)  \overset{n-1}{j}_r, \label{Frun}
\ee
for $n=0,1,\ldots$ with the understanding that $\overset{-1}{j}_r \equiv 0$. This allows us to express all $\overset{n}{F}_{ru}$ in terms of  $\overset{0}{F}_{ru}$ and the current. In fact, as shown in appendix \ref{maxwell-selfdual} $\overset{0}{F}_{ru}$  can in turn be expressed in terms of `free data' $\big( A_{A}(u,\hat{x}),\ \overset{0}{j}_{u}(u,\hat{x}) \big)$ as:
\be
\partial_u \overset{0}{F}_{ru} = \overset{0}{j}_{u} - 2 \partial_u D^{\zb} A_{\zb}. \label{fallofffru0}
\ee

\subsection{$u \to -\infty$ fall-offs and candidates for asymptotic charges}

As shown in \cite{stromprahar}, in the case of massless QED,  the asymptotic charge associated to large gauge transformations is given by\footnote{When $F$ is  self-dual, the charge is associated to negative helicity soft photon theorem.}

\begin{equation}
Q_{+}[\epsilon^{+}]\ =\ \int_{S^{2}}\epsilon^{+}(\hat{x})\ \overset{0}F_{ru}(u=-\infty,\hat{x})
\end{equation}

%Whence we see that assuming 
Now, the standard $u \to - \infty$ fall-offs for the radiative data  associated to generic solutions of Maxwell's equations is \cite{ash-stru}
\be
\overset{0}{F}_{ru}(u,\xh) = \overset{0,0}{F}_{ru}(\hat{x}) +O(u^{-\epsilon})
\ee
Thus the non-vanishing and finite limit of $\overset{0}{F}_{ru}$ at ${\cal I}^{+}_{-}$ is $\overset{0,0}{F}_{ru}(\hat{x})\ =\ \lim_{u\rightarrow -\infty}\overset{0}{F}_{ru}(u,\hat{x})$.

 Whence the charge density which gives rise to the leading (negative-helicity) soft theorem is given by 
 $\overset{0,0}{F}_{ru}(\xh)$.

Note that (\ref{Frun}) together with (\ref{fallofffru0}) implies $\overset{n}{F}_{ru}=O(u^n)+O(u^{n-\epsilon})$. However consider a subset of radiative fields with the following fall-off in $u$: 

\begin{equation}
\overset{0}{F}_{ru}(u,\xh) = \overset{0,0}{F}_{ru}(\hat{x}) +R(u), \label{fallFru0strong}
\end{equation}

where the remainder $R(u)$ falls off faster than $\vert u\vert^{-n}$ $\forall\ n$. 

Then using Eq. (\ref{Frun}), the  $u \to - \infty$ behavior of $\overset{n}{F}_{ru}$ can be specified to be
\be
 \overset{n}{F}_{ru}(u,\xh) = u^n \sum_{k=0}^n \frac{1}{u^k}  \overset{n,k}{F}_{ru}(\xh) + O(u^{-\epsilon})\ \forall\ n \label{uexp}
\ee 
One may think that this subspace is too restrictive to be physically interesting. However, as we show in appendix \ref{fallofftree}, these fall-offs are equivalent to the assumption that radiative data for gauge field $A_{A}(\omega,\hat{x})$ has a Laurent expansion in $\omega$ which begins at $\frac{1}{\omega}$. This precisely corresponds to the case of tree level scattering amplitudes where absence of infrared loop effects imply that Laurent expansion remains valid \cite{ashoke-biswajit}.

For this space of radiative data, there is a natural prescription to extract out a finite and non-trivial ``moment" of $\overset{n}{F}_{ru}$ as $u\ \rightarrow\ -\infty$.  From Eq. (\ref{uexp}) we see that $\overset{n,n}{F}_{ru}$ is the $u$-independent term of $\overset{n}{F}_{ru}$ as $u \to - \infty$.  This term may be thought of as an integration constant that arises when solving Eq. (\ref{Frun}). In terms of the lower order coefficients  $\overset{n,k}{F}_{ru}(\hat{x}), k<n$ this quantity can be expressed as:
\begin{equation}
\overset{n,n}{F}_{ru}(\hat{x})\ =\lim_{u\rightarrow -\infty}\left[\ \overset{n}{F}_{ru}(u,\hat{x})\ -\ \sum_{k=0}^{n-1}u^{n-k}\overset{n,k}{F}_{ru}(u,\hat{x})\ \right]
\end{equation}
As we show below, each of these $\overset{n,n}{F}_{ru}(\hat{x})$ generates an infinity of asymptotic charges which are in fact conserved in  classical theory so far as we restrict ourselves to the subspace of radiative data defined above.
We illustrate this idea with an example of charge whose Ward identities lead to subleading soft photon theorem for tree level scattering amplitudes.  That is, we will see how this charge (or equivalently fluxes which are integrals over ${\cal I}^{+}$) which was defined in \cite{stromlow, subleading} is nothing but
\begin{equation}
Q_{1}[\epsilon]\ =\ \int_{S^{2}}\epsilon(\hat{x})\ \overset{1,1}{F}_{ru}(\hat{x})
\end{equation}

\subsection{The subleading charge} \label{leq0F}

We would like to compute

\begin{equation}
\overset{1,1}{F}_{ru}(\hat{x})\ =\ \lim_{u\rightarrow\ -\infty}\left[\ \overset{1}{F}_{ru}(u,\hat{x})\ -\ u\overset{1,0}{F}_{ru}(\hat{x})\right]
\end{equation}

Now using Eq. (\ref{Frun}) we have, 
\be
2 \partial_u \overset{1}{F}_{ru} + \Delta  \overset{0}{F}_{ru}  =-2 D^z \overset{0}{j}_z + 2 \partial_u  \overset{0}{j}_r \label{Fru1}
\ee
In the $u \to -\infty$ limit the current term vanishes. Evaluating (\ref{Fru1}) with the expansion (\ref{uexp}) one finds
\be
2 \overset{1,0}{F}_{ru}+ \Delta \overset{0,0}{F}_{ru}=0.
\ee 
Thus, 
\be
\overset{1,1}{F}_{ru}(\xh) = \lim_{u \to - \infty} \left[\overset{1}{F}_{ru}(u,\xh) + \frac{u}{2} \Delta \overset{0}{F}_{ru}(u,\xh) \right] %=:\ \sigma_{(1)}(\hat{x})
\label{sigma0}
\ee
\begin{comment}
which cancels the linear in $u$ divergent term. This is the candidate subleading charge density. 
\end{comment}
To evaluate this ``charge density"  in terms of the free data, we write (\ref{sigma0})  as an integral over $u$ (here we  assume there is no contribution from $u=+\infty$, which is consistent with the absence of massive charges),
\ba
\overset{1,1}{F}_{ru}(\xh) & =&  - \int du \, (\partial_u \overset{1}{F}_{ru} + \frac{1}{2}\Delta \overset{0}{F}_{ru} + \frac{u}{2} \Delta \partial_u \overset{0}{F}_{ru}  ) \\
 & =& \int du \, ( D^z \overset{0}{j}_z  - \partial_u  \overset{0}{j}_r - \frac{u}{2} \Delta \partial_u \overset{0}{F}_{ru}) \label{sigma0final}
\ea  
where we used Eq. (\ref{Fru1}). The $\partial_u  \overset{0}{j}_r$, being a total $u$-derivative of a current term will not contribute and thus we have been able to express the charge-density as a flux (per unit solid angle) at ${\cal I}^{+}$ as
\begin{equation}
\overset{1,1}{F}_{ru}(\xh)\ =\   
\int du \, ( D^z \overset{0}{j}_z\ - \frac{u}{2} \Delta \partial_u \overset{0}{F}_{ru}) 
\end{equation}

If we now define $Q^{+}_{1}[\epsilon^{+}]$ as 

\ba
Q^{+}_{1}[\epsilon^{+}] & := & \int_{S^{2}}\epsilon^+(\hat{x})\overset{1,1}{F}_{ru}(\xh)\nonumber\\
&= & \int_{{\cal I}^{+}}\epsilon^{+}(\hat{x})\ ( D^z \overset{0}{j}_z\ - \frac{u}{2} \Delta \partial_u \overset{0}{F}_{ru} ) 
\ea
We immediately see that this charge matches with the charges obtained in \cite{subleading} which was shown to be associated to sub-leading soft photon (of negative helicity) insertions. In fact as shown in \cite{subleading}, $Q_{1}[\epsilon^{+}]$  equals the charge obtained in \cite{stromlow} if we define 
\begin{equation}
Y^{A}\ :=\ D^{A}\epsilon\ +\ \epsilon^{AB}D_{B}\epsilon
\end{equation}

An exactly analogous procedure yields the corresponding charge $Q_{1}^{-}[\epsilon^{-}]$ at ${\cal I}^{-}$. Along with anti-podal matching conditions on $\epsilon^{\pm}$, the conservation of charge would follow if we could show that
\begin{equation}
\overset{1,1}{F}_{ru}(\hat{x}) =  \overset{1,1}{F}_{rv}(-\hat{x}).
\end{equation}

We emphasize that in contrast to ideas and viewpoint employed in \cite{subleading}, here we never had to use covariant phase space techniques neither ``divergent" gauge transformations. Our analysis only refers to radiative data and structures available at null (and as we see below, spatial) infinity.

\subsection{Towards a tower of asymptotic charges}

Naturally we are now tempted to consider higher order ``charge densities"\footnote{Of course in order to justify calling them charge densities, we need to show that the corresponding charges are indeed conserved. This will be shown in section \ref{five}.} $\overset{n,n}{F}_{ru}(\hat{x})$ defined at ${\cal I}^{+}_{-}$.  As a first example of this kind, in this section we focus on $\overset{2,2}{F}_{ru}(\hat{x})$ and write the corresponding charge $Q^{+}_{2}[\epsilon^{+}]$ as an integral over ${\cal I}^{+}$ as a functional of the free data. We will refer to this charge as sub-subleading charge as the corresponding Ward identities will turn out to be equivalent to the (``projected'' \cite{llz}) sub-subleading soft photon theorem for tree-level QED.\footnote{The soft expansion at sub-subleading order does not factorize, and hence we have to project out the unfactorized contribution to get a sub-subleading soft photon theorem. We refer to these statements as projected soft theorems.}

That is we would like to evaluate,

\begin{equation}
\begin{array}{lll}
\overset{2,2}{F}_{ru}(\hat{x})\ =\ \lim_{u\rightarrow -\infty}\ \left[\ \overset{2}{F}_{ru}(u,\hat{x})\ -\ u^{2}\ \overset{2,0}{F}_{ru}(u,\hat{x})\ -\ u\overset{2,1}{F}_{ru}(u,\hat{x})\right]
\end{array}
\end{equation}

Using, Eq. (\ref{Frun}) we can immediately see that
\be
4 \partial_u \overset{2}{F}_{ru} + (\Delta + 2) \overset{1}{F}_{ru} = -2 D^z \overset{1}{j}_z + 2 \partial_u  \overset{1}{j}_r + 2  \overset{0}{j}_r, \label{Fru2}
\ee
Evaluating the equation in the $u \to - \infty$ and using (\ref{uexp}) as before one finds
\ba
8 \overset{2,0}{F}_{ru} + (\Delta +2) \overset{1,0}{F}_{ru} =0 \\
4 \overset{2,1}{F}_{ru}+(\Delta +2) \overset{1,1}{F}_{ru} =0 \label{Fru22}
\ea
From the first equation we find how to cancel the $u^2$ divergence of $\overset{2}{F}_{ru}$:
\ba
\overset{2}{F}_{ru} + \frac{u}{8}  (\Delta +2) \overset{1}{F}_{ru} & = & u( \overset{2,1}{F}_{ru}+ \frac{1}{8}  (\Delta +2) \overset{1,1}{F}_{ru} ) + O(1) \label{Fru21}\\
& = & - \frac{u}{8}(\Delta +2) \overset{1,1}{F}_{ru} +O(1), \label{Fru21b}
\ea
where in (\ref{Fru21}) we collected all  $O(u)$ terms and in (\ref{Fru21b})  we  simplified them using (\ref{Fru22}). Finally, from Eq. (\ref{sigma0}) we see that the $O(u)$ piece in (\ref{Fru21b}) can be cancelled by the addition of
\be
\frac{u}{8}(\Delta +2)\left[\overset{1}{F}_{ru} + \frac{u}{2} \Delta \overset{0}{F}_{ru} \right] .\label{Ou}
\ee
Adding (\ref{Ou}) to  (\ref{Fru21}) gives the candidate for the sub-subleading charge density
\be
\overset{2,2}{F}_{ru}(\xh) = \lim_{u \to - \infty}\left[\overset{2}{F}_{ru} + \frac{u}{4}  (\Delta +2) \overset{1}{F}_{ru} + \frac{u^2}{16} \Delta (\Delta+2) \overset{0}{F}_{ru} \right] .\label{sigma1}
\ee
As before, we now write (\ref{sigma1}) as an integral over $u$, assuming no contribution arises from  $u=+\infty$. Using (\ref{Fru2}) and (\ref{Fru1}), one again finds cancellation of terms, resulting in:
\be
\overset{2,2}{F}_{ru}(\xh)  =   \frac{1}{2}\int du \, \left[ D^z \overset{1}{j}_z - \partial_u \overset{1}{j}_r - \overset{0}{j}_r   + \frac{u}{2}(\Delta+2)(D^z \overset{0}{j}_z - \partial_u \overset{0}{j}_r) - \frac{u^2}{8}\Delta(\Delta+2) \partial_u \overset{0}{F}_{ru}\right] . \label{sigma1b}
\ee

The expression can be further simplified by integrating by parts the  term proportional to $u \partial_u \overset{u}{j}_r$ and dropping the total derivative term  $\partial_u \overset{1}{j}_r$. 

We note that unlike the leading and sub-leading charge densities (i.e. $\overset{0,0}{F}_{ru}(\hat{x})$ and $\overset{1,1}{F}_{ru}(\hat{x})$),  $\overset{2,2}{F}_{ru}(\hat{x})$ is a sum of two kinds of terms. First set of terms involve the integrand which is a local function of the free data, $\phi(u,\hat{x}), A_{A}(u,\hat{x})$ and in the second case, the integrand is a non-local function of the free data involving $\int^{u}\phi(u^{\prime},\hat{x})\ du^{\prime}$.

\ba
\overset{2,2}{F}_{ru}^{\text{non-local}} &:=&  \frac{1}{4}\int du \, [ 2 D^z \overset{1}{j}_z  + \Delta \overset{0}{j}_r  ]\\
\overset{2,2}{F}_{ru}^{\text{local}} &:=& \frac{1}{4}\int du \, \left[u (\Delta+2)D^z \overset{0}{j}_z - \frac{u^2}{4}\Delta(\Delta+2) \partial_u \overset{0}{F}_{ru} \right].
\ea
\begin{comment}
Whereas the `local' piece can be written entirely in terms of the free data $\phi(u,\xh), A_A(u,\xh)$ and its derivatives, the `non-local' piece involves terms with $\sint^u \phi(u,\xh)$.
\end{comment}

Using the above charge densities, we can define the corresponding charges, which are parametrized by   functions on the sphere %``large gauge parameters" 
$\epsilon(\hat{x})$ as, 
\ba
Q^+_{2}[\epsilon^{+}]\ =\ \int d^{2}x\ \epsilon^+(\hat{x})\overset{2,2}{F}_{ru}(\hat{x})\nonumber\\
Q^-_{2}[\epsilon^{-}]\ =\ \int d^{2}x\ \epsilon^-(\hat{x})\overset{2,2}{F}_{rv}(\hat{x})   \label{s2charge}
\ea
with $\epsilon^+(\xh)=  \epsilon^-(-\xh)=\epsilon(\xh)$. In fact, one can define a tower of asymptotic charges labelled by $n\ \geq 0$ and $\epsilon_n \in C^\infty(S^2)$:
\ba
Q^+_{n}[\epsilon^{+}_n]\ =\ \int d^{2}x\ \epsilon^+_n(\hat{x})\overset{n,n}{F}_{ru}(\hat{x})\nonumber\\
Q^-_{n}[\epsilon^{-}_n]\ =\ \int d^{2}x\ \epsilon^-_n(\hat{x})\overset{n,n}{F}_{rv}(\hat{x}), \label{sncharge}
\ea
with $\epsilon^+_n(\xh)=  \epsilon^-_n(-\xh)=\epsilon_n(\xh)$. We postpone to section \ref{four} a detailed description of these charges for arbitrary $n$.  For now we would like to comment on a particular feature of these charges that we will discover below: The spherical harmonic decomposition of $\overset{n,n}{F}_{ru}(\hat{x})$ starts at $l=n$. In other words, if we take $\epsilon_n = Y_{l,m}$, the charge is non-zero only for $l \geq n$. It turns out that the $l=n-1$ case corresponds to the so called Newman-Penrose charges \cite{npcharges}.

\subsection{Relationship with Newman Penrose charges} \label{NPsec}

We thus see we have a ``doubly infinite'' family of charges parametrized by $\left(n,\epsilon_n \in\ C^{\infty}(S^{2})\right)$ as in Eq. (\ref{sncharge}). As shown above and later in section \ref{four}, these charges (which are localised at $u = -\infty$ or $v=+\infty$) can be written as fluxes integrated over entire null infinity, if we assume that the radiative fields are trivial at $v= -\infty$ and $u = +\infty$. 

In \cite{npcharges}, Newman and Penrose showed that in a free theory with massless fields (that is pure Maxwell theory in our case),  one could construct infinitely  many ``charges" defined by taking a celestial 2-sphere located at $u =\textrm{constant}$ (or analogously, $v= \textrm{constant}$) and integrating certain densities which were constructed out of radiative data  and spherical harmonics $Y_{l,m}$. We note here that in the absence of sources, these charges are precisely an (infinite) subset of the asymptotic charges constructed above.\footnote{See \cite{condemao,condemao2,pope} for  earlier  explorations between soft theorems charges and NP charges.} That is if we consider the subset of charges parametrized by $\left(n+1, Y_{n,m}\right)$, then in vacuum these are the Newman Penrose (NP) charges.

NP charges are defined as follows (our choice of normalization is for later convenience)

\begin{equation}\label{NP1}
Q_{n}^{\textrm{NP}}\ = \frac{2}{(n+1)} \int d^{2}\hat{x}\ D^{z}Y_{n,m}(\hat{x})\ \overset{n}{F}_{rz}(u,\hat{x})
\end{equation}
Although the charge density is evaluated at a fixed $u$, by using vacuum equations of motion 
\begin{equation}\label{NP2}
2\partial_{u}\overset{n}{F}_{rz}\ =\ -\left[\frac{2}{n}D_{z}D^{z}\ +\ (n+1)\right]\overset{n-1}{F}_{rz}
\end{equation}
we can easily see that the charge is independent of $u$. Substituting Eq. (\ref{NP2}) in Eq. (\ref{NP1}) and integrating over the sphere, we see that due to defining equation of the spherical harmonics we have

\begin{equation}
\frac{d}{d u}Q_{n}^{\textrm{NP}}\ =\ 0.
\end{equation}

We can now see how these charges are related to the asymptotic charges we are considering in this paper. As our charges are in terms of $F_{ru}$ instead of $F_{rz}$ we need to find a relation between the two. This follows from the equation of motion involving the self-dual field,

\be
-\partial_{r} F_{rz}\ +2 \partial_{u} F_{rz}\ =\ D_{z}F^+_{ru}
\ee
Asymptotic expansion at  future null infinity  yields,

\be\label{NP3}
\overset{n+1}{F^+}_{ru}\ =\ -\frac{2}{ (n+1)}\ D^{z}\overset{n}{F}_{rz}.
\ee

Using Eq.(\ref{NP3}) in Eq.(\ref{NP1}) we see that NP charge is given by

\be
Q_{n}^{\textrm{NP}}\ =\ \int d^{2}\hat{x} Y_{n,m} \overset{n+1}{F^+}_{ru}(u,\hat{x}).
\ee

As these charges are finite, they can be evaluated at any $u$ and in particular at $u\ =\ -\infty$ they can be written as

\be
Q_{n}^{\textrm{NP}}\ =\  \int d^{2}\hat{x} Y_{n,m} \overset{n+1,n+1}{F^+}_{ru}(\hat{x}),
\ee
which are same as the asymptotic charges $Q_{n+1}[\epsilon]$ for $\epsilon\ =\ Y_{n,m}\ \forall\ m$. 

We conclude by noting  that  our fall-off conditions imply the vanishing of  the NP charges. This corresponds to the observation in \cite{npcharges} that the charges (at future null infinity) vanish for `outgoing solutions' (see beginning of p. 184 in \cite{npcharges}). The vanishing of NP charges can also be seen as a consequence of the relation with the soft theorem charges: In section \ref{four} we will see that the asymptotic charges can be written as
\be
\int d^{2}\hat{x} \, \epsilon \overset{n+1,n+1}{F}_{ru} = \int d^2 \xh du  (D^z)^{n+1}\epsilon \,  \rho_{z \ldots z}
\ee
where $\rho_{z \ldots z}$ depends on the free data at null infinity. Since $(D^z)^{n+1} Y_{n,m}=0$ this implies the vanishing of the NP charges.
\section{From Ward identities to sub-$n$ soft theorems}  \label{three}

We would now like to show that if we assume  $Q_{2}^{+}[\epsilon^{+}]\ =\ Q_{2}^{-}[\epsilon^{-}]$ then the corresponding Ward identity is equivalent to the sub-subleading soft photon theorem  in tree level scattering amplitudes \cite{shiu,llz}. 

However before establishing this equivalence we take a small detour and review the hierarchy of subleading soft theorems in tree level QED. At the 1st and 2nd order of hierarchy, these are nothing but sub-leading and sub-subleading soft photon theorems. As we will see, the higher order soft theorems are far less constraining then the previous ones, but are present nonetheless. We will refer to this entire hierarchy as sub-$n$ ($n \geq 1$) soft theorems

\subsection{sub-$n$ soft theorems}\label{3.1}

Consider an \emph{un-stripped} (that is, including the momentum conserving delta function) tree-level amplitude in QED consisting of $N$-charged particles and a photon of energy $\w$.\footnote{The sub-$n$ soft theorems were derived for stripped amplitudes in \cite{shiu,llz}. However it can be easily checked that the same factorization holds for unstripped amplitudes as well.}  Regarded as a function of $\w$, the amplitude has an expansion of the form
\be
\M_{N+1} = \sum_{n=0}^{\infty} \w^{n-1} \M^{(n)}_{N+1}. \label{gralexp}
\ee
As is well known, the first term is given by Weinberg's soft theorem\footnote{To simplify expressions we display signs as if all particles are outgoing and have positive charge. For negative outgoing or positive incoming charges the factor comes with opposite sign.}
\be
\M^{(0)}_{N+1} = \sum_{i=1}^N S^{(0)}(q,p_i) \M_N ,
\ee
\be
S^{(0)}(q,p) = e \frac{\e \cdot p}{q \cdot p}
\ee
where $\e^\mu$ is the photon polarization and  $q=(1,\qh)$ the photon 4-momentum direction. 

The next term  in the soft expansion also factorizes and is given by Low's subleading soft theorem:
\be
\M^{(1)}_{N+1} = \sum_{i=1}^N S^{(1)}(q,p_i) \M_N ,
\ee
\be
S^{(1)}(q,p) = \frac{e}{q \cdot p} \e_\mu q_\nu J^{\mu \nu}
\ee
where  $J^{\mu \nu} = p^\mu \partial_\nu - p^\nu \partial_\mu$ is the angular momentum of the particle with momentum $p$.\footnote{The sub-leading soft photon theorem is not universal even for tree-level scattering amplitudes. In fact it was shown in \cite{elvang} that there is a class of higher derivative terms which can modify the sub-leading factor. Thus the most general sub-leading factor comprises of the universal term  $S^{(0)}$ given above and an additive non-universal term. In this paper we restrict ourselves to universal sub-leading factor.}
For tree level scattering amplitudes and if we restrict to minimal coupling, in fact more is true. In \cite{llz} it was shown that gauge invariance implies the $n$-th term in (\ref{gralexp}) for $n \geq 1$ is given by
\be
\M^{(n)}_{N+1} = \frac{1}{n!}\sum_{i=1}^N S^{(1)}(q,p_i) (q \cdot \partial_i)^{n-1} \M_N  + \e_\mu q_{\nu_1} \ldots q_{\nu_{n-1}} A^{\mu \nu_1 \ldots \nu_{n-1}}(p_1,\ldots,p_N)\label{Ml}
\ee
where  $A^{\mu \nu_1 \ldots \nu_{n-1}}$ is antisymmetric under the exchange of $\mu$ and a $\nu$ index but its dependence on the hard momenta is undetermined by the requirement of gauge invariance. 

Thus at the onset it appears that there is no factorization theorem at sub-subleading order and beyond in tree-level QED. However as was shown in \cite{shiu}, the second term in Eq. (\ref{Ml}) can be projected out and one obtains, \emph{what we call} sub-n ($n \geq 1$) soft theorem. This can be understood as follows. 
\begin{comment}
In order to answer this question,  we will like to Follow the strategy for leading and sub-leading soft theorems and  like to construct a ``charge'' for  each such sub-$l$ factorization theorem. 
\end{comment}
The $n$-th term in (\ref{gralexp}) can be extracted as
\be
\lim_{\w \to 0} \partial^{n}_\w ( \w \M_{N+1} ) = \sum_{i=1}^N S^{(0)}(q,p_i) (q \cdot \partial_i)^{n-1} \M_N  + R^{(n)} \label{sublst}
\ee
where  $R^{(n)}$ denotes the reminder term in (\ref{Ml})
\be
R^{(n)} := \e_\mu q_{\nu_1} \ldots q_{\nu_{n-1}} A^{\mu \nu_1 \ldots \nu_{n-1}}. \label{Rl}
\ee

For definitiveness in the following  we restrict attention to the case where the  soft photon has  negative helicity. %\footnote{We remind the reader that  leading and sub-leading negative (positive) helicity soft photon theorems are equivalent to Ward identities of corresponding charges defined for self-dual (anti-selfdual)  fields respectively.}
Parametrizing the soft momentum direction  $q^\mu$ in terms of standard stereographic coordinates $q^\mu =(1,\hat{q}(w,\overline{w}))$ such that  $\e^{- \mu}= \frac{1}{\sqrt{2}}(w,1,i,-w)$  \cite{stromprahar} it can be easily seen that
\begin{equation}
D_{w}^{2} q_{\mu} = 0, \quad  \quad D^2_w  \big[ (1+ |w|^2)^{-1}\e^{-}_\mu \big]  =  0 ,
\end{equation}
as a result of which we have 

\begin{equation}
D^{n+1}_w  \big[ (1+ |w|^2)^{-1}R^{(n)} \big] =0.
\end{equation}

Thus by operating on both sides of Eq. (\ref{sublst}) with $D^{n+1}_w  (1+ |w|^2)^{-1}$ we get a factorization theorem which we call sub$-n$ soft theorem $\forall\ n \geq 1$:
\begin{equation}\label{sublst1}
\lim_{\w \to 0} \partial^{n}_\w ( \w D^{n+1}_w  (1+ |w|^2)^{-1}\M_{N+1} ) = D^{n+1}_w  (1+ |w|^2)^{-1}\left(\sum_{i=1}^N S^{(0)}(q,p_i) (q \cdot \partial_i)^{n-1}\right) \M_N .
\end{equation}

\subsection{Ward identities from soft theorems} \label{sec3.2}
In fact, from Eq. (\ref{sublst1}), we can write down the Ward identities which are equivalent to these soft theorems. Here we follow the same strategy that is used in deriving Ward identities from soft theorems in the leading and sub-leading case. 

We first note that the LHS of (\ref{sublst1}) corresponds to insertion of an operator 
 \be
\lim_{\w \to 0}\ D^{n+1}_{w}  (1+ |w|^2)^{-1}\ \partial^{n}_\w [\w a_-(\w \qh)] \label{unsmeared}
\ee
where  $a_-(\w \qh)$ is the Fock operator of a negative helicity photon. In analogy with other soft theorems we can construct a smeared version of (\ref{unsmeared}) such that it takes a simple form at null infinity. Recall the `free-data' of the Maxwell field at null infinity is given by the angular components of the vector potential, $A_{A}(u,\xh)$. These are related to the Fock operators by
\be
A_{\wb}(\w,\qh) = \frac{1}{4 \pi i} \frac{\sqrt{2}}{(1+ |w|^2)}a_-(\w \qh) \label{fdfock}
\ee
where $A_{\wb}(\w,\qh)$ is the time-Fourier transform of $A_{\wb}(u,\qh)$. We can hence see that by smearing  both sides of Eq.(\ref{sublst1}) by $\int d^{2}w\ T^{w\dots w}$ we obtain a formal\footnote{Formal in the sense that we need to show that this identity arises from conservation laws.} Ward identity  with a `soft' charge given by

\begin{comment}
The structure of the known charges associated to the leading and subleading soft theorems suggests the following candidate for  soft charge,
\end{comment}
\ba
\Qs_n[T] & := & %\int du \, u^{l+1} \partial_u  \int d^2 \qh \, T  D^{l+1}_w D^{\wb} A_{\wb}(u,\qh) \label{softcharge1}\\
(-i)^{n+1} \lim_{\w \to 0}   \partial^{n}_\w  \, \w \int d^2 w \, T D^{n+1}_w A_{\wb}(\w,\qh) \label{softcharge2}
\ea
where 
\be
T \equiv T^{\overbrace{w \ldots w}^{n}}
\ee
is the holomorphic component of a rank $n$ (symmetric, trace-free) sphere tensor that parametrizes the charge.   In going from the first to second line we  used that $\sqrt{\gamma} \gamma^{w \wb}=1$.  
From (\ref{fdfock}) and (\ref{softcharge2}), the soft charge can be written as the operation 
\be
 \frac{(-i)^{n}}{4 \pi}  \int d^2 w \, T  D^{n+1}_w  \frac{\sqrt{2}}{(1+ |w|^2)}   \label{smearing}
\ee
acting on  (\ref{unsmeared}).  In other words, the smearing (\ref{smearing}) acting on the LHS of the sub-$n$ soft theorem (\ref{sublst}) can be interpreted as arising from the insertion of a soft charge (\ref{softcharge2}). In order  to look for the existence of a  hard charge, we need to perform the  same smearing  on the RHS of (\ref{sublst1}). 
\begin{comment}
The first observation is that under this smearing, the reminder term $R^{(l)}$ disappears, namely:
\be
D^{l+2}_w  \big[ (1+ |w|^2)^{-1}R^{(l)} \big] =0.
\ee
This result can be shown using the identities,
\ba
 D^2_w q_\mu &= &0,  \label{D2q}\\
D^2_w  \big[ (1+ |w|^2)^{-1}\e^{-}_\mu \big] & = & 0 
\ea
and the fact that there are $l+2$ derivatives and $l$ powers of $q_\mu$ in $R^{(l)}$.
We now discuss the smearing (\ref{smearing}) on the first term of  the RHS of (\ref{sublst}). 
\end{comment}

Up to an $i/e$ factor, such smearing defines the following differential operator on the $p$ variable,
\be
\T :=  \frac{(-i)^{n+1}}{2}  \int d^2 w \, T  D^{n+1}_w \left[ \Kwb  (q \cdot \partial)^{n-1} \right], \label{deltaT}
\ee
\be
\text{where} \quad \Kwb  :=\frac{1}{2\pi e} \frac{\sqrt{2}}{(1+ |w|^2)}  S^{(0)}_-(q,p) 
\ee

(the minus subscript denotes the negative helicity of the soft photon under consideration). 
We will see this operator satisfies  two key properties:
\begin{enumerate}
\item $\T$ is local in the $p$ variable
\item $\T$ satisfies
\be
\int \dpp b^\dagger  (\T b)  = (-1)^{n} \int \dpp  (\T b^\dagger) b .  \label{sympl}
\ee
\end{enumerate}
The first property will be a consequence of the identity \cite{stromlow}:
\be
D^2_w \Kwb  = D_w \delta^{(2)}(w,z) \partial_E + E^{-1} \delta^{(2)}(w,z) \partial_z, \label{D2K}
\ee
where $(E,z,\zb)$ parametrize the momentum $p$ of the hard particle.  The second property will ensure the smearing (\ref{smearing}) on the RHS of (\ref{sublst}) can be understood  as arising from a hard charge:\footnote{The $(-1)^n$ sign arises because, in the factorization formula, an incoming 4-momentum is expressed as outgoing by reversing its sign.} 
\ba
[b,\Qh_n[T]]  & = & i e \T b,  \label{commbQ} \\  \label{commbQdag}
[b^\dagger,\Qh_n[T]]  &= & i e (-1)^{n} \T b^\dagger. 
\ea
 Classically,  Eq. (\ref{sympl}) is the condition that the infinitesimal transformation $\delta b = \T b$, $ \delta b^* =  (-1)^{n} \T b^*$ is symplectic. The charge reproducing (\ref{commbQ}) and (\ref{commbQdag}) will then be  given by:
 \be
 \Qh_n[T] = \frac{i e}{2 (2 \pi)^3} \int_0^\infty d E E \int d^2 \xh \, b^\dagger(E,\xh) \T b(E,\xh)  - (b \leftrightarrow c), \label{Qhard}
 \ee
where $(b \leftrightarrow c)$ is the contribution from the antiparticles.

To compare with the usual definition of hard charges which are integrals over ${\cal I}$, we will finally need to write (\ref{Qhard}) in terms of the free data of the scalar field at ${\cal I}$, $\phi(u,\xh)$. Equivalently we can write the hard charge as integral over $(E,z,\zb)$ where $E$ is the energy conjugate to $u$ (or $v$). Whence if $\phi(E,\xh)$ and $\bar{\phi}(E,\xh)$ denote the Fourier transforms of $\phi(u,\xh)$ and $\bar{\phi}(u,\xh)$, one has
\be
\begin{array}{lll} 
\phi(E,\xh) = \frac{b(E \xh)}{4 \pi i} , & \quad \bar{\phi}(E,\xh) = \frac{ c(E \xh) }{4 \pi i},  & \quad \text{for} \quad E>0 \\
& & \\
\phi(E,\xh) = - \frac{c^\dagger(-E \xh)}{4 \pi i} , &  \quad \bar{\phi}(E,\xh) = -\frac{ b^\dagger(-E \xh) }{4 \pi i}, & \quad \text{for} \quad E<0
\end{array}
\ee
From these expressions, and using the fact that under $E \to -E$, $\T \to (-1)^{n+1} \T$, the charge (\ref{Qhard}) can be written as
\be
\Qh_n[T] = \frac{i e}{2 \pi} \int_{-\infty}^\infty d E E \int d^2 \xh \, \bar{\phi}(-E,\xh) \T \phi(E,\xh)  - (\phi \leftrightarrow \bar{\phi}). \label{Qhard2}
\ee
 We now calculate $\T$ and  associated charges $Q_{n}[T] = \Qs_{n}[T] + \Qh_{n}[T]$ for $n=1,2$. We expect these charges to be related to the subleading and sub-subleading charges $Q_{n}[\epsilon],  n=1,2$ which were defined in the previous section.

%We now review how the above results specialize to the known charges associated to the sub-leading soft theorem \cite{strominger-low, campi-low}.

\subsection{$n=1$ case: subleading  charge from soft theorem}
When $n=1$ one finds
\be
\T= \frac{1}{2}(D_z T \partial_E - T E^{-1}\partial_z).
\ee
It is easy to verify
\be
\int_0^\infty dE E \int d^2 \xh ( b^\dagger \T b + \T b^\dagger b) = 0,
\ee
which corresponds to property (\ref{sympl}).  To compute the charge (say at ${\cal I}^{+}$) we use the identities
\ba
\int \frac{dE}{2\pi} E \bar{\phi}(-E)  \partial_E \phi(E) & = & \int du u \partial_u \bar{\phi}(u)  \phi(u)\\
\int \frac{dE}{2\pi}\bar{\phi}(-E)  \phi(E) &  = & \int du  \bar{\phi}(u)  \phi(u) .
\ea
One can then verify (\ref{Qhard2}) becomes
\be
\Qh_1[T]= \frac{1}{2} \int du d^2 \xh( D_z T^z u \overset{0}{j}_u +T^z \overset{0}{j}_z) .
\ee
On the other hand, the soft charge (\ref{softcharge2}) can be written as
\be
\Qs_1[T] = -\int du d^2 \xh  D_z T^z u \partial_u D^{\zb} A_{\zb}.
\ee
Finally, taking $T^z = - 2 D^z \epsilon$ with $\epsilon(\xh)$ a function on the sphere one can verify  the total (hard plus soft) charge takes the form
\be
Q_1[T^z= - 2 D^z\epsilon] = \int d^2 \xh \epsilon(\xh) \overset{1,1}{F}_{ru}(\xh)
\ee
with the charge density $\overset{1,1}{F}_{ru}(\xh)$ found in section \ref{leq0F}  (Eqs. (\ref{sigma0final}) and (\ref{fallofffru0})). 
\subsection{$n=2$ case: sub-subleading charge from soft theorem}
For $n=2$ one finds that, 
\be
\T= \frac{i}{2} \left[ D^2_z T \partial_E^2 -2 E^{-1} D_z T \partial_E \partial_z +2  E^{-2}(D_z T \partial_z +T D_z \partial_z ) \right] .\label{Tl1}
\ee
One can check it verifies condition (\ref{sympl}). To compute the charge we  use the identities
\ba
\int \frac{dE}{2\pi} E^{-1} \bar{\phi}(-E)  \phi(E) & = & i \int du (\sint^u \bar{\phi})  \phi(u)\\
\int \frac{dE}{2\pi} \bar{\phi}(-E)   \partial_E  \phi(E) & =& i \int du u \bar{\phi}(u)  \phi(u)  \\
\int \frac{dE}{2\pi} E \bar{\phi}(-E)  \partial_E^2  \phi(E) &  = & i \int du  u^2 \partial_u \bar{\phi}(u)  \phi(u) .
\ea
One then finds (\ref{Qhard2}) to be given by
\be
\Qh_2[T] = -\frac{1}{2}\int du d^2 \xh \left[  u^2 D^2_z T^{zz}  \overset{0}{j}_u +2  u D_z T^{zz}  \overset{0}{j}_z + 2 i e [ (\sint^u \bar{\phi}) (D_z T^{zz}\partial_z \phi +T^{zz}D^2_z \phi) - (\phi \leftrightarrow \bar{\phi})] \right] .\label{Qhard1}
\ee
Thus hard charge is a sum of two terms. One term is an integral over local functionals of the radiative data and the other term involves non-local fields which are then integrated. In the above equations, these are the terms which involve $\int^{u}\bar{\phi}$ (or $\int^{u}\phi$).

On the other hand, the soft charge (\ref{softcharge2}) can be written as
\be
\Qs_2[T] = \int du d^2 \xh  D^2_z T^{zz} u^2 \partial_u D^{\zb} A_{\zb}.
\ee
To compare with  $\overset{2,2}{F}_{ru}(\xh)$ obtained in the previous section, we take
\be
T^{zz}= \frac{1}{2} D^z D^z \epsilon.
\ee
Using the identity  $[D_z,D^z] V^z = V^z$ one verifies that the sum of local terms in $\Qh_2[T]$ added to  $\Qs_2[T]$ yield the `local' part of $\overset{2,2}{F}_{ru}$,
\be
Q^{\text{local}}_2[T^{zz}= \frac{1}{2} D^z D^z \epsilon] = \int d^2 \xh \epsilon(\xh)  \overset{2,2}{F}_{ru}^{\text{local}}(\xh).
\ee
As shown in subsection \ref{non-loc}, there is a similar matching between the  `non-local' terms,
so that the total charges coincide:
\be
Q_2[T^{zz}= \frac{1}{2} D^z D^z \epsilon] = \int d^2 \xh \epsilon(\xh)   \overset{2,2}{F}_{ru}(\xh). \label{match1}
\ee

Whence the question we would like to ask is if  the charges $Q_{n}[T]$ defined from the  sub-$n$ soft theorems are also the same as $Q_{n}[\epsilon]$ defined  at ${\cal I}$ when $n\ >\ 2$. We turn to this question in  section \ref{four}.
\subsubsection{Matching  the non-local terms}\label{non-loc}

We want to show the matching of the  `non-local' terms in Eq. (\ref{match1}). This corresponds to the equality
\be
\int du  d^2 \xh \, \epsilon(\xh)  ( 2 D^z \overset{1}{j}_z  + \Delta \overset{0}{j}_r ) = -  i e \int du d^2 \xh  \left[ (\sint^u \bar{\phi}) (D_z T^{zz}D_z \phi +T^{zz}D^2_z \phi) - (\phi \leftrightarrow \bar{\phi}) \right] \label{nl0}
\ee
with $T^{zz} \equiv \frac{1}{2} D^z D^z \epsilon$. We start with the integrand of the LHS of (\ref{nl0}) using the expressions for the current in terms of the scalar field given in Eq. (\ref{l1j}). Discarding total sphere-derivative terms we have: (we omit a  ``$- (\phi \leftrightarrow \bar{\phi})$'' at the end of each equation)
\ba
 \epsilon ( 2 D^z \overset{1}{j}_z  + \Delta \overset{0}{j}_r )  & = & i e [ - 2 D^z \epsilon ( \phi D_z  \overset{1}{\bar{\varphi}} -  \overset{1}{\bar{\varphi}} D_z \phi) - \Delta \epsilon  \phi  \overset{1}{\bar{\varphi}} ]   \\
 & = & 4 i e  \overset{1}{\bar{\varphi}} D^z \epsilon D_z \phi \\
 & = & - 2 i e (\sint^u \bar{\phi})  \Delta( D^z \epsilon D_z \phi)  \\
 & = & -  i e (\sint^u \bar{\phi})[2 D_z T^{zz} D_z \phi + 2  T^{zz}D^2_z \phi+ 2 D^z \epsilon D_z \Delta \phi + \Delta \epsilon \Delta \phi ] ,\label{nl4}
\ea 
where we used  $\Delta f = 2 D_z D^z f = 2 D^z D_z f$ for a scalar function $f$. The last two terms in (\ref{nl4}) can be brought to a  different form by integrating by parts in $u$. Up to total $u$ and sphere derivatives one can show the identity
\be
(\sint^u \bar{\phi}) [2   D^A \phi D_A f +  \phi \Delta f ]- (\phi \leftrightarrow \bar{\phi})=0 \label{idl1b}
\ee
for any $u$-independent sphere function $f(\xh)$. Using this identity for $f= \Delta \epsilon$, as well as the sphere-derivative relations 
\be
[D^z,D_z]V_z = V_z, \quad [D_z,D^z]V^z = V^z
\ee
one can show:
\be
(\sint^u \bar{\phi})[ 2 D^z \epsilon D_z \Delta \phi + \Delta \epsilon \Delta \phi +   D_z T^{zz} D_z \phi +   T^{zz}D^2_z \phi ] -  (\phi \leftrightarrow \bar{\phi})=0. \label{idl1}
\ee
Using (\ref{idl1}) in (\ref{nl4}) one arrives at the desired result in (\ref{nl0}).

\section{Higher order charges and a conjectured equivalence} \label{conj-equi} \label{four}
\def\rhons{\overset{n}{\rho}_{\text{soft}}}
\def\rhonh{\overset{n}{\rho}_{\text{hard}}}

In this section we extend our previous discussion to the higher order soft theorems. That is, we would like to argue that sub-$n$ soft theorems for $n\ >\ 2$ are equivalent to Ward identities associated to $Q_{n}[\epsilon]$. %\footnote{To simplify notation we  omit a  label `$n$' in the smearing function $\epsilon$.} 
Although we do not provide a complete proof, we give several hints in this direction and conjecture the equivalence between sub-$n$ soft theorems and Ward identities of higher-$n$ charges. 

Before embarking on a tedious analysis, let us summarize what we are able to prove below. We show that if one defines a relationship between $T$ and $\epsilon$ as,

\begin{equation}
T\ :=\ 2\frac{(-1)^{n}}{n!^{2}}(D^{z})^{n}\epsilon
\end{equation}
then 
\begin{equation}
Q_{n}[\epsilon]^{\textrm{soft}}\ =\ Q_{n}[T]^{\textrm{soft}}
\end{equation}
For the hard charges, we are  able to bring $Q_{n}[\epsilon]^{\textrm{hard}}$ and $Q_{n}[T]^{\textrm{hard}}$ into a form that allows for a direct comparison. Establishing their equality however becomes a technical problem we have not been able to solve for arbitrary $n$.

The first step in understanding the relationship between $Q_{n}[T]$ and $Q_{n}[\epsilon]$ is to write $Q_{n}[\epsilon]$ in terms of free data at ${\cal I}^{\pm}$. As always we focus on analyzing the charge at ${\cal I}^{+}$.

%We then propose their equivalence with the tower of soft theorem charges described in section 4. 
 In order to simplify expressions, let us introduce the following notation:
\be \label{defDeltan}
\Delta_n := -\frac{1}{2n}(\Delta +n(n-1)), \quad n \geq 1 ,
\ee
\be \label{defDeltanm}
\Delta(n,m) := \prod_{k=m}^n \Delta_k, \quad 1 \leq m \leq n ; \quad \Delta(n,0) := 0, \quad \Delta(n,n+1):=1
\ee
\be
s_n:= \frac{1}{2n}( -2 D^z \overset{n-1}{j}_z + 2 \partial_u  \overset{n-1}{j}_r + n  \overset{n-2}{j}_r)  , \quad n \geq 1 ,
\ee
where as before  $\overset{-1}{j}_r \equiv 0$.
The field equations (\ref{Frun}) can then be written as:
\be
 \partial_u \overset{n}{F}_{ru} = \Delta_n  \overset{n-1}{F}_{ru} +s_{n} , \quad n \geq 1 \label{feq}.
\ee

\subsection{$n$-th asymptotic charge in terms of free data}

Consider  the quantity 
\be
\sigma_n(u,\xh) :=  \sum_{m=0}^n \frac{(-u)^{n-m}}{(n-m)!} \Delta(n,m+1)  \overset{m}{F}_{ru}(u,\xh). \label{sigman}
\ee
Using the field equations (\ref{feq}) it is straightforward to show that
\be
\partial_u \sigma_n = \frac{(-u)^n}{n!}\Delta(n,1)\partial_u \overset{0}{F}_{ru}+  \sum_{m=1}^{n} \frac{(-u)^{n-m}}{(n-m)!} \Delta(n,m+1) s_{m} .  \label{prop2}
\ee
Furthermore,  the  $u \to -\infty$ fall-off conditions described in section 2.1 together with the field equations (\ref{feq}) can be shown to imply (see appendix \ref{prop1app}): 
\be
\lim_{u \to -\infty} \sigma_n(u,\xh)= \overset{n,n}{F}_{ru}(\xh) \label{prop1} .
\ee

Properties (\ref{prop2}) and (\ref{prop1}) can be used  to express $\overset{n,n}{F}_{ru}(\xh)$ in terms of data at null infinity by the same  procedure as in the $n \leq 2$ charges studied before. We start with Eq. (\ref{prop1}) and an integration by parts in $u$:
\be
\overset{n,n}{F}_{ru}(\xh)  =  \lim_{u \to -\infty} \sigma_n(u,\xh) = - \int du \, \partial_u \sigma_n(u,\xh) + \lim_{u \to \infty} \sigma_n(u,\xh).
\ee
In the absence of massive fields the last term vanishes (this term should otherwise  give the contribution to the charge from the massive particles) and one is left with the integral over $u$. From Eq.  (\ref{prop2}) and  Eq. (\ref{fallofffru0}), 
\be
\partial_u \overset{0}{F}_{ru} = \overset{0}{j}_{u} - 2 \partial_u D^{\zb} A_{\zb}, \label{feq0}
\ee
one can write the resulting expression as a sum of `soft' and `hard' pieces,
\be
\overset{n,n}{F}_{ru}(\xh) = \int  du \big( \rhons(u,\xh) + \rhonh(u,\xh) \big)
\ee
where
\be
 \rhons : =   \frac{2 (-u)^n}{n!} \Delta(n,1) \partial_u D^{\zb}A_{\zb} 
 \ee
\be
\rhonh : = -\frac{(-u)^n}{n!} \Delta(n,1) \overset{0}{j}_u +  \sum_{m=1}^{n} \frac{(-u)^{n-m}}{(n-m)! m} \Delta(n,m+1) \big(  D^z \overset{m-1}{j}_z -  \partial_u  \overset{m-1}{j}_r - \frac{1}{2} m  \overset{m-2}{j}_r \big). \label{rhonh}
\ee
We finally need to  express the current coefficients in terms of  the scalar field radiative data $\phi$. For the tree-level charges of interest here the  scalar field can be treated as free when computing the current. For the first term of (\ref{rhonh}) we have
\be
\overset{0}{j}_u = i e \phi \partial_u \bar{\phi} + c.c.
\ee
For the second term, one can show  the combination of current coefficients  can be written as
\be\label{phik1}
[D^z \overset{m-1}{j}_z +\frac{1}{2(m-1)}\Delta \overset{m-2}{j}_r] = \frac{2 i e}{(m-1)} \sum_{k=1}^{m-1} k D^z( \overset{k}{\varphi} D_z \overset{m-1-k}{\bar{\varphi}} )   - (\varphi \leftrightarrow \bar{\varphi})
\ee
where $ \overset{k}{\varphi}(u,\xh)$ is the $1/r^{k+1}$ coefficient in the $1/r$ expansion of the  scalar field $\varphi$. These coefficient can be expressed in terms of the free data by recursively solving  the free field equation
\be
\partial_u  \overset{n}{\varphi} = \Delta_n \overset{n-1}{\varphi} \implies  \overset{k}{\varphi} = \Delta(k,1) \partial_u^{-k} \phi, \label{phik}
\ee
where $\partial_u^{-k}$ denotes the $k$-th $u$-primitive. Substituting (\ref{phik}) in (\ref{phik1}) and then in (\ref{rhonh}) gives an expression of $\rhonh(u,\xh)$ in terms of the scalar field free data $\phi(u,\xh)$.

\subsection{Higher-n charges from soft theorems}
From the discussion of section \ref{sec3.2}, the (smeared) sub-$n$ soft theorem can be written as
\be
Q_n[T] S = S Q_n[T]
\ee
where $Q_n[T] = \Qs_n[T] + \Qh_n[T]$ with $\Qs_n[T]$ and $\Qh_n[T]$ given in Eqs. (\ref{softcharge2}) and (\ref{Qhard2}) respectively. 
We now discuss how to cast these quantities in terms of free data at null infinity. For the soft charge, we simply Fourier transform (\ref{softcharge2}) to obtain
\be
\Qs_n[T]  =   \int du d^2 \qh \,  u^{n}     T  D^{n}_w \partial_u D^{\wb} A_{\wb}(u,\qh) .
\ee
For the hard charge we need to bring the differential operator $\T$ defined in Eq. (\ref{deltaT}) into a simpler form. As in the $n=1,2$ cases, the integral localizes in the hard momentum direction. The general form, derived in appendix  \ref{Tapp}, is found to be
\be
\T = - \frac{i^{n+1} n!}{2}\sum_{k=0}^{n}   \frac{(-1)^k }{(n-k)!}  D^{n-k}_z T \partial^{n-k}_E E^{-k} D_z^k . \label{Tfinal}
\ee
Substituting this in the  expression for the hard charge (\ref{Qhard2}) and using the Fourier transform identities
\be
\int \frac{d E}{2 \pi} E \bar{\phi}(-E,\xh)  \partial^{n-k}_E E^{-k} D_z^k  \phi(E,\xh) = - i^{n+1-2k} \int du u^{n-k} \partial_u \bar{\phi}(u,\xh) \partial_u^{-k}  D_z^k \phi(u,\xh)
\ee
one arrives at 
\be
 \Qh_n[T]=   \frac{i e}{2}  (-1)^{n+1} n! \sum_{k=0}^n \frac{1}{(n-k)!}\int du d^2 \xh \; u^{n-k} D^{n-k}_z T  \partial_u \bar{\phi} \partial_u^{-k}  D_z^k \phi - (\phi \leftrightarrow \bar{\phi}) \label{Qhard3}
\ee

\subsection{Conjectured equivalence}
We wish to generalize the equivalence found in section \ref{three} between  $\overset{n,n}{F}_{ru}$  and the soft theorem charges  $Q_n[T]$. The idea is to parametrize the tensor $T$ in terms of a sphere function $\epsilon(\xh)$ as $T \sim (D^z) \epsilon$, and then identify $\epsilon(\xh)$ with a smearing function for  $\overset{n,n}{F}_{ru}(\xh)$. That is, we wish to find $T_\epsilon$ such that
\be
\int d^2 \xh   \epsilon(\xh) \overset{n,n}{F}_{ru}(\xh) = Q_n[T_\epsilon]. \label{comparison}
\ee
Using the identity
\be
(D_z)^n (D^z)^n \epsilon = (-1)^n n! \Delta(n,1) \epsilon \label{Dzn}
\ee
one can show  the \emph{soft} part of (\ref{comparison}) is satisfied provided one sets
\be
T_\epsilon := \frac{2 (-1)^n}{n!^2} (D^z)^n \epsilon. \label{Teps}
\ee
Thus, in order to establish  (\ref{comparison}), one needs to show the matching (\ref{comparison}) between the \emph{hard} parts,
\be
 Q_{n}[\epsilon]^{\textrm{hard}}  = \Qh_n[T_\epsilon] \label{hardcomparison}.
\ee
with $T_\epsilon$ given by (\ref{Teps}). %As discussed below, establishing (\ref{hardcomparison}) boils down to a combinatorial problem we unfortunately have not been able to solve. 

A  strategy to compare both sides of (\ref{hardcomparison}) is to bring the RHS into a form where $\epsilon$ appears with no derivatives. Substituting (\ref{Teps}) in (\ref{Qhard3}) and using the identity\footnote{To show (\ref{Dznk}), apply $D^k_z$ on both sides of the equation.  The LHS is then given by  (\ref{Dzn}). For the RHS use (\ref{Dzn}) for $n=k$ and $\Delta(k,1) \Delta(n,k+1)=\Delta(n,1)$.}
\be
(D_z)^{n-k} (D^z)^n \epsilon =  \frac{(-1)^{n} n!}{(-1)^{k} k!} (D^z)^k \Delta(n,k+1) \epsilon, \quad k=0,\ldots, n \label{Dznk}
\ee
one finds, after integration by parts on the sphere,
\be
\Qh_n[T_\epsilon] =  i e   (-1)^{n+1} \sum_{k=0}^n \frac{ 1}{k! (n-k)!} \int du d^2 \xh \; u^{n-k} \epsilon(\xh) \Delta(n,k+1)  (D^z)^k[ \partial_u \bar{\phi} \partial_u^{-k}  D_z^k \phi ]- (\phi \leftrightarrow \bar{\phi}). \label{Qhard4}
\ee
Expression (\ref{Qhard4}) defines a ``charge density'' that should  be compared with $\rhonh(u,\xh)$ as given in Eqs. (\ref{rhonh}) to (\ref{phik}). One can readily check that the $k=0$ term in (\ref{Qhard4}) agrees with the first term in (\ref{rhonh}). The comparison of the remaining $k \geq 1$ terms in  (\ref{Qhard4}) with the second term in (\ref{rhonh}) is non-trivial due to the possibility of integration by parts in $u$ and the  interchanges $\varphi \leftrightarrow \bar{\varphi}$. This was already seen in the $n=1,2$ cases studied earlier, where the equality of charges required the use of several non-trivial identities. Unfortunately, we have not been able to find a general form of such  identities that would allow us to establish the equality for arbitrary $n$.

% This  should then be compared with the $k \geq 1$ terms in  (\ref{Qhard4}). As already seen for the $n=1,2$ charges, the comparison is . We unfortunately have not been able to find the general structure so as to establish the equivalence for $n\ >\ 2$. 
%(although we have obtained some partial results, as for instance identifying in (\ref{Qhard4}) the $m=1,2$ terms in (\ref{rhonh})).

\section{Proof of conservation laws in classical theory} \label{class-cons} \label{five}

In the previous sections we showed that the sub-$n$ soft theorems can be written as Ward identities of a tower of asymptotic charges and conjectured that these charges were the charges generated from $\overset{n,n}{F}_{ru}(\hat{x})$ at ${\cal I}^{+}$ and $\overset{n,n}{F}_{rv}(\hat{x})$ at ${\cal I}^{-}$. Our conjecture was motivated by showing this equivalence in the case of leading, sub-leading and sub-subleading soft photon theorems.

However even assuming the conjecture to be valid, a natural question arises. Why do we expect this infinite hierarchy of infinite dimensional (asymptotic) charges to be conserved in classical theory? After all, the theory is not expected to be integrable.
 In this section we show that classically these charges are indeed conserved. That is, by analyzing the theory at \emph{spatial infinity}, we show that\footnote{In this section ${F}_{ab}$ will stand for the \emph{real} field strength rather than its self-dual part. The Ward identity associated to the negative (positive) sub-$n$ soft theorem corresponds  to the (anti) self-dual part of Eq. (\ref{cons}). The conservation statements can be understood as Eq. (\ref{cons}) for the real field strength, together with its Hodge dual version.}

\begin{equation}\label{cons}
\overset{n,n}{F}_{ru}(\hat{x}) = \overset{n,n}{F}_{rv}(-\hat{x}).
\end{equation}
Our proof is an extension of the analysis done in \cite{eyhe} for the $n=0$ case.   We will summarize the key idea first and then provide a detailed analysis of these conservation laws.

\subsection{Key ideas} \label{5.1}

$\overset{n,n}{F}_{ru}(\hat{x})$ and $\overset{n,n}{F}_{rv}(\hat{x})$ are charge densities localised on $S^{2}$ at $u = -\infty$ and $v = +\infty$ respectively. These two celestial spheres can also be understood as (future and past) boundaries of a three dimensional de Sitter space representing space-like infinity of Minkowski (or more generally asymptotically flat) spacetime in a  particular compactification \cite{romano}. More in detail, when working with hyperbolic coordinates $(\tau,\rho,z,\zb)$, if we take $\rho\rightarrow\ \infty$ while keeping $\tau,z,\zb$ fixed, we reach Lorentzian three dimensional de Sitter. We will refer to this (blow up of) spatial infinity as $\H$.  The advantage of working with this definition of spatial infinity is that it dovetails nicely with ${\cal I}^{\pm}$. Thus the boundary spheres at $\tau\ =\ \pm\infty$ are mapped onto $S^{2}$ at $u = -\infty$, $v = \infty$ respectively. 

In order to prove Eq. (\ref{cons}), we need to relate the fields at $\H$ with fields on the respective boundaries of ${\cal I}^{\pm}$ . By analyzing the equations of motion at spatial infinity and assuming appropriate fall-offs of the fields as $u \rightarrow -\infty$ and $v \rightarrow \infty$ we relate fields at $\H$ with $\overset{n,n}{F}_{ru}(\hat{x})$ and $\overset{n,n}{F}_{rv}(\hat{x})$.

We will thus first analyze the equations of motion at spatial infinity in section \ref{5.2}  and in section \ref{5.3}, show how to relate $\overset{n,n}{F}_{ru}(\hat{x})$ and $\overset{n,n}{F}_{rv}(\hat{x})$ with certain  data at $\H$.

\subsection{Maxwell equations at spatial infinity}\label{5.2}
We introduce hyperbolic coordinates in the region $r>|t|$,
\be
x^\mu = \rho Y^\mu(y), \quad Y^\mu(y) Y_\mu(y) = 1, 
\ee
where $\rho:= \sqrt{x^\mu x_\mu} = \sqrt{r^2 -t^2}$ and $y = y^\alpha$  are coordinates on the unit hyperboloid $\H$. The Minkowski line element takes the form
\be
ds^2 =  d \rho^2 + \rho^2 h_{\alpha \beta} d y^\alpha d y^\beta,
\ee
with $h_{\alpha \beta}$ the unit hyperboloid metric. For concreteness we take $y^\alpha= (\t,x^A)$, $\t=t/\sqrt{r^2-t^2}$ as coordinates on $\H$ so that
\be
Y^\mu = ( \tau, \sqrt{1+\t^2} \; \xh),
\ee
\be
h_{\alpha \beta} d y^\alpha d y^\beta= - \frac{d \tau^2}{1+\tau^2}+(1+\tau^2) q_{AB} dx^A dx^B .
\ee
Following \cite{beig}, we assume a $1/\rho$ expansion of the Maxwell field near spatial infinity:
\ba
F_{ \rho \alpha}(\rho,y) & =& \frac{1}{\rho} \sum_{n=0}^{\infty} \frac{1}{\rho^n} \overset{n}{F}_{\rho \alpha }(y) \label{Frhoalphaexp}\\
F_{\alpha \beta}(\rho,y) & =&  \sum_{n=0}^{\infty} \frac{1}{\rho^n} \overset{n}{F}_{\alpha \beta}(y) .
\ea
For the scalar field, we assume that their fall-off behavior at spatial infinity is faster than $O(\rho^{-m})\ \forall\ m$. This assumption is motivated by the fact that we do not consider soft limit of the charged particles.  Under this assumption the field equations reduce to source-free Maxwell equations at spatial infinity. The reason we can consistently do this is due to the fact that there is a clear separation between the source and electromagnetic radiation. In the case of non-abelian gauge theories or perturbative gravity at second order there is no such separation and hence our analysis has to be generalized. Note that for massive charged particles the assumption can be proved since in a flat background massive fields decay exponentially fast at spatial infinity.

Under this expansion, the source-free Maxwell equations take the form\footnote{We remind the reader that in this section we are  working with real (rather than self-dual) field strengths.}
\be
\D^\alpha \overset{n}{F}_{\rho \alpha} =0, \quad \D^\beta \overset{n}{F}_{\alpha \beta} + n \overset{n}{F}_{\rho \alpha} =0 \label{divFspi}
\ee
\be
\partial_{[\alpha} \overset{n}{F}_{\beta \gamma]}=0, \quad \partial_\alpha \overset{n}{F}_{\rho \beta} - \partial_\beta \overset{n}{F}_{\rho \alpha}+n \overset{n}{F}_{\alpha \beta}=0,  \label{bianchispi}
\ee
where $\D_\alpha$ is the covariant derivative on $\H$.   For $n=0$ the equations and their solutions are related to the leading soft photon theorem charges \cite{eyhe}. The idea is that the subleading charges will be related to the remaining values of $n$. For $n>1$  one can decouple the equations by combining the second equations of (\ref{divFspi}), (\ref{bianchispi}) to eliminate  $\overset{n}{F}_{\alpha \beta}$. Using the identity $[\D^\beta,\D_\alpha] V_\beta = 2 V_\alpha$ one arrives at \cite{perng}:
\be
\D^\alpha \overset{n}{F}_{\rho \alpha} =0, \quad [\D^2 +(n^2-2)] \overset{n}{F}_{\rho \alpha} =0,  \quad n>0. \label{Espi}
\ee
Once a solution to $\overset{n}{F}_{\rho \alpha}$ is found, one can obtain $\overset{n}{F}_{\alpha \beta}$ from the second equation in  (\ref{bianchispi}).  Our next task is to specify $\t \to \infty$ fall-offs. Due to the divergence-free condition, the $\alpha=A$ and $\alpha=\t$ fall-offs are not independent: If we assume  $\overset{n}{F}_{\rho A} =O(\t^h)$ then $\overset{n}{F}_{\rho \t} =  O(\t^{h-3})$.  When such asymptotics are inserted in the wave equation (\ref{Espi}) one finds   $h= \pm n$. In the next subsection we show that our prescribed fall-offs at null infinity imply $h=-n$. Thus, at spatial infinity the problem to solve is  (\ref{Espi}) with the $\t \to \infty$ asymptotic condition
\be
\overset{n}{F}_{\rho A}(\t,\xh) = \t^{-n} \overset{n,0}{F}_{\rho A}(\xh)+ \ldots. \label{fallFtau}
\ee
The leading sphere components $\overset{n,0}{F}_{\rho A}(\xh)$ play the role of `asymptotic free data' for the  hyperboloid vector field $\overset{n}{F}_{\rho \alpha}$. On the other hand, the leading term of the $\alpha=\t$ component,
\be
\overset{n}{F}_{\rho \t}(\t,\xh) = \t^{-n-3} \overset{n,0}{F}_{\rho \t}(\xh)+ \ldots. \label{Frhotaum0}
\ee
is determined by the asymptotic divergence-free condition,
\be\label{rhoalph-rhoA}
n \overset{n,0}{F}_{\rho \t} + D^A \overset{n,0}{F}_{\rho A} =0.
\ee

\subsection{Relating field expansions %Equations of Motion 
at  null and spatial infinity} \label{5.3}

As shown in appendix \ref{fallofftree} in the limit $u \to -\infty$, the falloff for the $1/r^{k+2}$ coefficient of $F_{ru}$ is given by, 
\be
 \overset{k}{F}_{ru}(u,\xh) = u^k \sum_{l=0}^k \frac{1}{u^l}  \overset{k,l}{F}_{ru} + \ldots. \label{uexp2}
\ee 
where the dots denote terms that fall off faster than any power of $1/u$. We now use this expansion to obtain the $\t \to \infty$ fall-offs at spatial infinity.

Combining (\ref{rexpFru}) with (\ref{uexp2}) we obtain the double sum expansion\footnote{In this section $\overset{k,l}{F}_{ru}$  denotes the coefficients of the real field strength  $F_{ru}$  rather than  its self-dual part $F^+_{ru}$. The two have the same  fall-off behavior.}
\be
F_{ru}(r,u,\xh)= \sum^\infty_{k=0}  \sum_{l=0}^{k} (u/r)^{k+2} (1/u)^{l+2}\overset{k,l}{F}_{ru}(\xh). \label{doublesumnull}
\ee
We regard this as an expansion in the two small parameters:
\be
|u/r| \ll 1 , \quad \text{and} \quad |1/u| \ll 1, \label{smallnull}
\ee
%which intuitively means that we are  taking $r \to \infty$ faster than $u \to - \infty$. 
For later comparison with the spatial infinity expansion, it will be convenient to rewrite (\ref{doublesumnull}) in a way that the $l$-sum appears first:
 \be
 F_{r u}(r,u,\xh) =  \sum_{l=0}^\infty \sum_{k=l}^{\infty}  (1/u)^{l+2}  (u/r)^{k+2}\overset{k,l}{F}_{ru}(\xh). \label{doublesumnull2}
 \ee

We now express the two small parameters (\ref{doublesumnull2}) in terms of the $(\rho,\t)$ coordinates. From the change of coordinates,
\be
r = \rho \sqrt{1+\t^2}, \quad u = \rho(\t - \sqrt{1+\t^2}),
\ee
one finds:
\be
1/u= - \frac{2 \t}{\rho} \big(1+O(\t^{-2}) \big),
\ee
\be
u/r= -\frac{1}{2\t^2}+O(\t^{-4}).
\ee
Substituting these in (\ref{doublesumnull2}) yields an expansion in the small parameters
\be
\t/\rho \ll 1 , \quad \text{and} \quad 1/\t^2 \ll 1, \label{smallspi}
\ee
that can be related to the spatial infinity expansion.  Note that the $\rho$ dependence appears only in the $l$-sum. From the relation 
\be
F_{\rho \t}= \frac{\rho}{\sqrt{1+\t^2}} F_{ru}
\ee
one finds  the $\rho$ dependence is the same as in the expansion (\ref{Frhoalphaexp}):
\be
F_{\rho \t}(\rho,\t,\xh)  = \frac{1}{\rho} \sum_{l=0}^{\infty} \frac{1}{\rho^l} \overset{l}{F}_{\rho \t }(\t,\xh).
\ee
Furthermore, for a given $1/\rho$ power, the dominant $\t\to \infty$ term is determined by the  $k=l$ summand. Specifically, one finds:
\be
\overset{l}{F}_{\rho \t}(\t,\xh) = \t^{-l-3} \overset{l,l}{F}_{ru}(\xh)+ O( \t^{-l-5}),
\ee
as anticipated in Eq. (\ref{Frhotaum0}). We thus conclude that
\be \label{spinull1}
 \overset{n,0}{F^+}_{\rho \t}(\xh)= \overset{n,n}{F}_{ru}(\xh),
\ee
where we included a  $+$ upperscript  to indicate this is a $\tau \rightarrow + \infty$ coefficient\footnote{Not to be confused with the self-dual field. We apologize for the overlap of notation.}.  The analogous  analysis at $\tau \rightarrow -\infty$ yields

\begin{equation} \label{spinull1v}
\overset{n,0}{F^-}_{\rho\tau}(\hat{x})\ =\ (-1)^{n+1}\ \overset{n,n}{F}_{rv}(\hat{x})
\end{equation}

A similar analysis for the other components of the field strength (in the limit $\tau\ \rightarrow\ \pm\infty$)  revels that
\be \label{spinull2}
 \overset{n,0}{F}_{\rho A}(\xh) =  \overset{n-1,n}{F}_{r A}(\xh)
 \ee
where $\overset{n-1,n}{F}_{r A}$ is the $O(u^0)$ coefficient of $\overset{n-1}{F}_{r A}$. Eqs. (\ref{spinull1}), (\ref{spinull2}) (and the analogues for past infinity) are the key equations that will allow us to relate the future and past charge densities.  Notice that relation  (\ref{rhoalph-rhoA}) becomes, upon using (\ref{spinull1}), (\ref{spinull2}),  equivalent to  relation (\ref{NP3}) discussed in section \ref{NPsec}.\footnote{Recall in our conventions $F_{ab}= \frac{1}{2}(F^+_{ab}+F^-_{ab})$ and $F^-_{rz}=0=F^+_{r\zb}$. Adding (\ref{NP3}) to its complex conjugated then leads to Eq. (\ref{rhoalph-rhoA}).}

%This is an important equation. In conjunction with Eq. (\ref{rhoalph-rhoA}), it tells us how to determine $\overset{n,n}{F}_{ru}$ (and analogously $\overset{n,n}{F}_{rv}$) in terms of the ``free data'' $\overset{n,0}{F}_{\rho A}$ at $\tau\ \rightarrow\ \infty$ at $i^{0}$.

\subsection{Establishing the conservation} \label{six}
From the perspective of spatial infinity it is natural to work with $\overset{n,0}{F}_{\rho A}$ ($\equiv \overset{n-1,n}{F}_{r A}$) as the charge density, since it represents `free data' for the differential equation (\ref{Espi}). The idea is to start with data at the future asymptotic boundary of $\H$ and evolve it backwards to the asymptotic past  of $\H$. One may object that this  conservation is rather trivial: One is simply casting the free field equations in hyperbolic coordinates, but for free fields the conservation should be trivial! The non-triviality arises from the strong fall-off conditions imposed at future and past null infinities. As shown in the previous subsection,  these select $O(|\t|^{-n})$ fall-offs on $\H$ which need not be satisfied for generic solutions to Eq. (\ref{Espi}).  It turns out that  this specific large $\t$ decay is related to another consequence of the strong fall-offs at null infinity:  that the (vector) spherical harmonic decomposition of ($\overset{n-1,n}{F}_{r A}$) $\overset{n,n}{F}_{ru}(\xh)$ starts at $l=n$. % (and analogous statement for  $\overset{m-1,m}{F}_{r A}(\xh)$), see section \ref{two}). 
As shown below,   the $O(|\t|^{-n})$ fall-off together with the $l\geq n$ spherical harmonic property imply that solutions to Eq. (\ref{Espi}) satisfy specific parity conditions on $\H$ under inversion $Y^\mu \to - Y^\mu$. This parity property then automatically implies the conservation law.

Our strategy is to find an explicit `boundary to bulk' Green's function that solves (\ref{Espi}) and satisfies all the aforementioned properties. A first candidate can be found by generalizing the Green's functions used to extend superrotations vector fields to time-infinity \cite{massivebms}. This leads to the following family of Green's functions:
\be \label{candidate0}
G_A^\alpha(\qh,y) \sim  (Y \cdot q)^{-(n+2)} J^{\mu \nu}_\alpha(Y) L_{\mu \nu}^B(q),
\ee
where $J^{\mu \nu}_\alpha$ and $L_{\mu \nu}^B$ are the angular momentum vector fields on $\H$ and $S^2$ respectively, 
\be
J^{\mu \nu}_\alpha(Y) = Y^\mu \D_\alpha Y^\nu - (\mu \leftrightarrow \nu),
\ee
\be
L_{\mu \nu}^A(q) = q_\mu D^A q_\nu - (\mu \leftrightarrow \nu).
\ee
With the help of identities given in appendix \ref{idsH} one can readily verify that (\ref{candidate0}) satisfies Eq. (\ref{Espi}) (with respect to the $y$ variable). It is also not difficult to verify that solutions  constructed from (\ref{candidate0}) have the `wrong' $O(\t^n)$ fall-offs.\footnote{We expect the Green's functions (\ref{candidate0}) can be used to define  \emph{smearing fields}   which would allow to  extend  the  charges $Q_n[\epsilon]$ to spatial infinity, as in the $n=0$ case \cite{eyhe}. We leave for future work the study of smeared charges at spatial infinity. See  also \cite{perng} for a set of closely related charges.}

To look for the Green's function with the `correct' $O(\t^{-n})$ fall-offs, we seek to replace  $(Y \cdot q)^{-(n+2)}$ in (\ref{candidate0}) by another  homogenous function  $f(Y \cdot q)$. There are two independent homogenous functions $f(s)$ such that   $f(\lambda s)= \lambda^{-{(n+2)}} f(s)$: $f(s) \propto s^{-(n+2)}$ and $f(s) \propto \delta^{(n+1)}(s)$ (the $(n+1)$-th derivative of the Dirac delta function). We thus consider the following ansatz for the solution to Eq. (\ref{Espi}):\footnote{The ansatz (\ref{candidate}) was also inspired by the integral representation of solutions to free Maxwell equations used by Herdegen, see e.g. \cite{herdegen}.}
\be
\overset{n}{F}_{\rho \alpha}(\t,\xh) = \frac{1}{2} \int d^2 \qh \, \delta^{(n+1)}(Y \cdot q) J^{\mu \nu}_\alpha(y) L_{\mu \nu}^B(\qh) \overset{n}{V}_B(\qh), \label{candidate}
\ee
where  $\overset{n}{V}_A(\qh)$ is an arbitrary sphere vector field. With the help of the identities given in appendix \ref{idsH} (together with the Dirac delta identities $s^2 \delta^{(n+2)}(s) = (n+2) (n+1) \delta^{(n)}(s)$ and  $s \delta^{(n+1)}(s) = - (n+1) \delta^{(n)}(s)$) one can verify that (\ref{candidate})  indeed satisfies  (\ref{Espi}). Studying the $\t \to \infty$ fall-off of (\ref{candidate}) is more subtle and we leave it to appendix \ref{fallt}. We find there that (\ref{candidate}) has the correct fall-offs (\ref{fallFtau}), (\ref{Frhotaum0}) with\footnote{The operator $ \Delta(n,1) $ acting on covectors is given as in the definition for scalars,  Eqs. (\ref{defDeltan}), (\ref{defDeltanm}) but with $\Delta$ replaced by $\Delta -1$ so that $D^A \Delta(n,1) V_A =  \Delta(n,1) D^A V_A$.}
\be
\overset{n,0}{F}_{\rho A}  = -  n 2  \pi (-1)^{n+1}  \Delta(n,1) \overset{n}{V}_A \label{fallt1}
\ee
\be
\overset{n,0}{F}_{\rho \t} = 2 \pi (-1)^{n+1} \Delta(n,1) D^B \overset{n}{V}_B.  \label{fallt2}
\ee
These equations give the relation between our actual data at the asymptotic future of $\H$ with the auxiliary field $\overset{n}{V}_A$. To write the solution in terms of our data we need to invert these relations, i.e. express $\overset{n}{V}_A$ in terms of $\overset{n,0}{F}_{\rho A}$. This can be done provided $\overset{n,0}{F}_{\rho A}$ has a vector spherical harmonic expansion starting at $l=n$ (corresponding to a spherical harmonic expansion of $\overset{n,0}{F}_{\rho \t}$ starting at $l=n$). This is precisely the property that follows from the equivalence with the soft theorem charges (also responsible for the vanishing of the NP charges, see section \ref{NPsec}). %We thus explicitly see the compatibility between the $l \geq n$ property with the $O(\t^{-n})$ fall-off and 
We thus conclude that (\ref{candidate}) gives the correct solution to our problem. Since (\ref{candidate}) is even under $Y^\mu \to -Y^\mu$, we have that the leading $\t \to \pm \infty$ coefficients of $\overset{n}{F}_{\rho \t}$ satisfy
\be \label{consspi}
\overset{n,0}{F^+}_{\rho \t}(\xh) = (-1)^{n+1} \overset{n,0}{F^-}_{\rho \t}(-\xh) 
\ee
 where the $\pm$ upperscripts  distinguish the coefficients from  the $\t \to + \infty$ and $\t \to - \infty$ expansions. This relation, together with Eqs. (\ref{spinull1}) and (\ref{spinull1v}), implies (\ref{cons}).

%\section{Conservation of charges in classical theory: A proof from spatial infinity} \label{six}

\section{Conclusions and outlook}

In this paper we have shown that for tree level scattering processes, there exists an infinite hierarchy of conservation laws such that at each level $n$ of the hierarchy  there is an infinite dimensional family of conserved charges $Q_{n}[\epsilon_n]$  labeled by functions on the sphere $\epsilon_n(\xh)$.\footnote{For $\epsilon_n=Y_{l,m}$ (a  spherical harmonic of order  $l$)  the charge vanishes if  $l < n$. For $l=n-1$ the expression coincides with the $(n-1)$-th Newman-Penrose charge, but it vanishes due to our assumed $|u| \to \infty$ fall-offs.} For $n=\ 0,1$ these charges are the well known asymptotic charges associated to leading and sub-leading soft photon theorems. Our analysis divorced the derivation of asymptotic charges from the paradigm of large gauge transformations. In fact, as  argued in \cite{subleading}, we do not expect the charges corresponding to levels $n\ \geq 2$ to be associated to any large gauge transformations. These are intrinsically boundary charges and are purely a consequence of the fall-off behavior of the radiative data and equations of motion of the theory at infinity.  A natural question is, what kind of factorization theorems arise by considering Ward identities associated to charges with $n\ \geq\ 2$. 

We showed that the  Ward identities associated to $n=2$ charges are equivalent to the sub-subleading soft photon theorem and conjectured that this equivalence holds $\forall\ n$. That is, Ward identities associated to $Q_{n}[\epsilon_n]$ are equivalent to sub-$n$ soft  photon theorems. Note that these sub-$n$ soft theorems only capture the factorized piece of the $\w^{n-1}$ coefficient  in the $\w \to 0$ expansion of the amplitude. For  $n \geq 2$ there are non-factorized terms that are not seen by these charges. 

Beyond the obvious open issue of proving the aforementioned conjecture, there are several open questions which emerge out of this work and which have not been addressed in this paper. We expect that our analysis can easily be generalized to tree-level amplitudes in perturbative gravity. In fact, an interesting puzzle in this direction concerns the soft-exactness of tree level MHV amplitudes in pure gravity \cite{huang}. That is, in the case of Gravity MHV amplitudes, the sub-$n$ soft limit exactly factorizes and there is no remainder $R_{n}$. Whence we expect the asymptotic charges to constrain the MHV amplitudes completely as opposed to only in the infrared sector. 
%Hence it will be interesting to explore the integrability properties of (the extension of) asymptotic charges in gravity. 

A related question concerns the algebra  of  charges $Q_{n}[\epsilon_n]$. If it does form an algebra then the quantization of this infinite dimensional algebra could shed interesting light in the structure of gauge theories. %(very vague remark).

Once loop corrections are taken into account, the story changes completely beyond the leading order in soft limit. As shown in \cite{ashoke-biswajit}, the sub-leading soft photon theorems receives corrections at one loop and are associated to terms proportional to $\ln\omega$ as opposed to $\omega^{0}$. Such a soft expansion clearly implies that the large $u$ fall-off behavior that we have assumed for tree level scattering data breaks down and one needs to consider a slower fall-off where the radiative field contains $\frac{1}{u}$ terms as $\vert u\vert\ \rightarrow\ \infty$  \cite{ashoke-memory}.  Whether such soft theorems with logarithmic corrections can be derived from asymptotic charges remains to be seen.

\section{Acknowledgements}

We are grateful to  Ashoke Sen for raising the issue of proving classical conservation law associated to subleading soft photon theorem which led us to this inquiry as well as for discussions on infinity of soft theorems. We are also grateful to Amitabh Virmani for urging us to look at the relationship between asymptotic charges and Newman-Penrose charges.  AL would like to thank IISER, Pune and   ICTS, Bangalore for their hospitality during course of this project.  MC would like to thank the  Center for Theoretical Physics at Columbia University  for  hospitality during the final stages of this project.

\appendix

\section{Maxwell equations for self-dual fields at ${\cal I}^{+}$}\label{maxwell-selfdual}

In retarded coordinates, the components of the 4-volume form take the form
\be
\e_{ruAB}= r^2 \e_{AB}
\ee
with $\e_{AB}$  the unit-sphere area form. The duality operation for the various components reads:
\be
\tilde{F}_{ru}= \frac{1}{2 r^2} \e^{AB}F_{AB}, \quad \tilde{F}_{AB} = -r^2 \e_{AB} F_{ru}, \label{Frudual}
\ee
\be
\tilde{F}_{rA}= \e_{A}^{\phantom{A} B}F_{rB}, \quad \tilde{F}_{uA} = \e_{A}^{\phantom{A} B}(F_{rB}-F_{uB}).
\ee
The idea  is to use Eqs. (\ref{Fpluseqs}) to arrive at an equation for $F^+_{ru}$ in terms of the source.  We then consider Eqs. (\ref{eqr}), (\ref{equ}), (\ref{eqA}) with $F_{ab}$ replaced by $F^+_{ab}$. From (\ref{eqA}) and the self-duality condition one arrives at:
\be
\e_{A}^{\phantom{A} B} j_B = - i \partial_r F^+_{uA} + i \partial_u F^+_{rA} - i D_A F^+_{ru}. \label{eqepsA}
\ee
Next we combine the sphere divergences of Eqs. (\ref{eqA}) and (\ref{eqepsA}) so as to keep only the $j_z$ part of the current. Since in holomorphic coordinates the area form is given by
\be
\e_{z \zb} = i \gamma_{z \zb},
\ee
the appropriate combination is:
\be
2 D^z j_z \equiv (q^{AB}- i \e^{AB}) D_A j_B = -\partial_r D^A F^+_{rA} + 2 \partial_u D^A F^+_{rA} - \Delta F^+_{ru}, \label{eqjz}
\ee
where we used that $D^A D^B F^+_{AB}=0$. Finally, using Eq. (\ref{eqr}) one can write  $F^+_{rA}$ appearing in (\ref{eqjz}) in terms of $j_r$ and $F^+_{ru}$. The resulting equation only involves $F^{+}_{ru}$ and the current components $j_z$ and $j_r$. Such equation can then be expanded in $1/r$ to yield  a recursive relation for the $F^+_{ru}$ components. Assuming the standard $1/r$ expansion,
\ba
F^+_{ru}(r,u,\xh) & = & \frac{1}{r^2} \sum_{n=0} \frac{1}{r^n}\overset{n}{F}_{ru}(u,\xh) \label{rexpFruapp}\\
j_z(r,u,\xh) & = & \frac{1}{r^2} \sum_{n=0} \frac{1}{r^n}\overset{n}{j}_z(u,\xh) \label{rexpjzapp} \\
j_{r}(r,u,\xh) & = & \frac{1}{r^4} \sum_{n=0} \frac{1}{r^n}\overset{n}{j}_{r}(u,\xh)  \label{rexpjrapp} 
\ea
(to simplify notation we omit the $+$ superscript in the coefficients for $F^+_{ru}$) one finds,
\be
2(n+1) \partial_u \overset{n+1}{F}_{ru} + (\Delta + n(n+1)) \overset{n}{F}_{ru} = -2 D^z \overset{n}{j}_z + 2 \partial_u  \overset{n}{j}_r + (n+1)  \overset{n-1}{j}_r, \label{Frunapp}
\ee
for $n=0,1,\ldots$ with the understanding that $\overset{-1}{j}_r \equiv 0$. This allows to express all $\overset{n}{F}_{ru}$ in terms of  $\overset{0}{F}_{ru}$ and the current. $\overset{0}{F}_{ru}$  can in turn be expressed in terms of `free data' by using the leading equations in (\ref{equ}) and (\ref{Frudual}):
\be
\partial_u \overset{0}{F}_{ru} = \overset{0}{j}_{u} - 2 \partial_u D^{\zb} A_{\zb}, \label{Fru0}
\ee
where $A_A(u,\xh)$ is the leading term of the sphere components of the vector potential and we assumed $A_u = O(r^{-1})$ so that $F_{u A}(r,u,\xh) = \partial_u A_A(u,\xh) +O(r^{-1})$.

In fact by using the leading equations in (\ref{equ}) and (\ref{Frudual}) it can be  easily shown that, 

\be
\partial_u \overset{0}{F}_{ru} = \overset{0}{j}_{u} - 2 \partial_u D^{\zb} A_{\zb}, \label{Fru0app}
\ee
where $A_A(u,\xh)$ is the leading term of the sphere components of the vector potential and we have assumed that $A_u = O(r^{-1})$ so that $F_{u A}(r,u,\xh) = \partial_u A_A(u,\xh) +O(r^{-1})$.

\section{$u \to \pm \infty$ fall-offs} \label{fallofftree}

The $\omega \to 0$ expansion of the tree-level amplitude (\ref{gralexp}) implies the radiative free data for the photon field has a similar expansion in frequency space:
\be
\tilde{A}(\w,\xh) = \w^{-1} \tilde{A}^{(-1)}(\xh)+ \tilde{A}^{(0)}(\xh)+\w A^{(1)}(\xh) + \ldots
\ee
In real space this translates into an  expansion of the type
\be
A(u,\xh) =\theta(u) A^{(\theta)}(\xh)+ \delta(u) A^{(\delta)}(\xh)+\delta'(u)A^{(\delta')}(\xh) + \ldots,
\ee
with $\theta$ the step function. Thus, the $u \to \pm \infty$ falloffs for the null infinity free data are of the type
\be
A(u,\xh) \overset{u \to \pm \infty}{=} A^{\pm}(\xh) + \ldots
\ee
where the dots are terms that go to zero faster than any power of $1/u$. For the matter field $\phi(u,\xh)$ we assume a similar expansion but with vanishing $u \to \pm \infty$ limit (corresponding to the fact that there are no `soft' charge particles).

Equations (\ref{Frun}) and (\ref{fallofffru0})  imply then the following $u \to -\infty$ falloff for the $1/r^{k+2}$ coefficient of $F_{ru}$:
\be
 \overset{k}{F}_{ru}(u,\xh) = u^k \sum_{l=0}^k \frac{1}{u^l}  \overset{k,l}{F}_{ru} + \ldots. \label{uexp2app}
\ee 
where the dots denote terms that fall-off faster than any power of $1/u$.

\section{Asymptotics of charged field and current at null infinity} \label{nulljapp}
For the massless scalar field we assume the standard $1/r$ expansion at null infinity:
\be
\varphi(r,u,\xh) = \frac{1}{r}\sum_{n=0}^\infty \frac{1}{r^n} \overset{n}{\varphi}(u,\xh). \label{rexpphi}
\ee
This implies a $1/r$ expansion for the current
\be
j_a = i e \varphi \partial_a \bar{\varphi} + c.c.,
\ee
as in Eqs. (\ref{rexpjzapp}), (\ref{rexpjrapp}) with coefficients
\ba
\overset{n}{j}_r & = & - i e \sum_{k=0}^{n+1} (n-2k+1) \overset{k}{\varphi} \; \overset{n-k+1}{\bar{\varphi}} \\
\overset{n}{j}_u & = & i e \sum_{k=0}^n \overset{k}{\varphi}\, \partial_u  \overset{n-k}{\bar{\varphi}} +c.c.\\
\overset{n}{j}_A & = & i e \sum_{k=0}^n \overset{k}{\varphi}\, \partial_A  \overset{n-k}{\bar{\varphi}} + c.c.
\ea
The actual current sourcing the field strength has an extra term  $-2 e^2 A_\mu \varphi \bar{\varphi}$. However such term is not seen in the tree-level formula we are interested here. 

The scalar field coefficients in the expansion (\ref{rexpphi}). Should be determined by the field equations. For the tree-level expression of interest here, it  suffices to consider the free-field equation $\square \varphi=0$, which takes the form
\be
\partial_u  \overset{n}{\varphi} +\frac{1}{2n}(\Delta +n (n-1))\overset{n-1}{\varphi}=0. \label{eqphi}
\ee
The lowest components of the current coefficients  are 
\be
\overset{0}{j}_u = i e \phi \partial_u \bar{\phi} + c.c., \quad \overset{0}{j}_A = i e \phi \partial_A \bar{\phi} + c.c. 
\ee
\be
\overset{0}{j}_r = - i e \phi \overset{1}{\bar{\varphi}} + c.c., \quad \overset{1}{j}_A = i e (\phi \partial_A \overset{1}{\bar{\varphi}} +\overset{1}{\bar{\varphi}} \partial_A \phi)+ c.c.  \label{l1j}
\ee
where $\phi \equiv \overset{0}{\varphi}$ is the scalar field `free data' and 
\be
 \overset{1}{\varphi}(u,\xh) = -\frac{1}{2} \Delta \int^u \phi(u',\xh) du'
\ee
according to (\ref{eqphi}).

\section{Appendix for section \ref{conj-equi}}

\subsection{Eq. (\ref{prop1})} \label{prop1app}

At $u \to -\infty$ we have the expansion (\ref{uexp}):
\be
\overset{m}{F}_{ru}(u,\xh) \overset{u \to -\infty}{=} \sum_{k=0}^m u^{m-k} \overset{m,k}{F}_{ru}(\xh) \label{Fuinf} .
\ee
Substituting  (\ref{Fuinf}) in (\ref{sigman}) and interchanging the order of the sums one arrives at
\be
\sigma_n \overset{u \to -\infty}{=} \overset{n,n}{F}_{ru}  + (-1)^n \sum_{k=0}^{n-1} u^{n-k}  \sum_{m=k}^{n} \frac{(-1)^{m}}{(n-m)!} \Delta(n,m+1) \overset{m,k}{F}_{ru}. \label{sigman2}
\ee
In order to establish (\ref{prop1}) we need to show that the $m$-sum in (\ref{sigman2}) vanishes for all $k=0,\ldots,n-1$.   The field equations (\ref{feq}) imply  the coefficients $\overset{m,k}{F}$ satisfy 
\be
\overset{m,k}{F}_{ru} = \frac{1}{(m-k)} \Delta_m \overset{m-1,k}{F}_{ru}, \quad  k<m
\ee
from which one finds
\be
\overset{m,k}{F}_{ru} = \frac{1}{(m-k)!} \Delta(m,k+1) \overset{k,k}{F}_{ru} \label{FrkFkk}.
\ee
Using (\ref{FrkFkk}), the $m$-sum in (\ref{sigman2}) takes the form
\ba
\sum_{m=k}^{n} \frac{(-1)^{m}}{(n-m)!} \Delta(n,m+1) \overset{m,k}{F}_{ru} & = & \sum_{m=k}^{n} \frac{(-1)^{m}}{(n-m)!(m-k)!} \Delta(n,k+1)\overset{k,k}{F}_{ru} \\
&=& 0
\ea
where we used that $\Delta(n,m+1) \Delta(m,k+1) = \Delta(n,k+1)$ and the fact that $\sum_{m=k}^{n} \frac{(-1)^{m}}{(n-m)!(m-k)!}=0$. This proves (\ref{prop1}).

\subsection{Eq. (\ref{Tfinal})} \label{Tapp}
%{ \miguel Need to rewrite with  $n \to n-1$ and include  multiplicative factor $ \frac{(-i)^{n+1}}{2}$ in expression for  $\T$ so that it coincides with  Eq. (\ref{deltaT}). With such replacements the final result of the appendix, Eq. (\ref{Tfinalapp}) should coincide with (\ref{Tfinal}).}

We start with the  expression for the differential operator 
\be
\T :=   \int d^2 w \, T  D^{n+2}_w \left[ \Kwb  (q \cdot \partial)^{n} \right], \label{TT}
\ee
where $T \equiv T^{\overbrace{w \ldots w}^{n+1}}$. The conventions in this appendix are slightly different than in the rest of the paper. The differential operator in the main text, Eq. (\ref{deltaT}), is obtained from (\ref{TT}) by doing the replacement $n \to n-1$ and including a multiplicative factor $ \frac{(-i)^{n+1}}{2}$. We will work here with the form given in (\ref{TT}) and make the above replacements at the end of the calculation.

We start by noting that $\partial_\mu  \equiv \frac{\partial}{\partial p^\mu}$ in (\ref{TT}) is an ``off-shell'' derivative. We parametrize this off-shell momentum and its derivative by:
\be
p^\mu = (p^0, \rho \ph)
\ee
\be
\partial_\mu = (\partial_0, \ph^i \partial_\rho + \rho^{-1} D^A \ph^i \partial_A), \label{partialmu}
\ee
with $\ph$  parametrized by $(z,\zb)$.  For the on-shell momentum we will have $p^0= \rho = E$ and
\be
 \partial_E = \partial_0 + \partial_\rho.
\ee
Using Blanchet-Damour multi-index notation:
\be
q^{N} \equiv q^{\mu_1} \ldots q^{\mu_{n}}, \quad  \partial_{N} \equiv \partial_{\mu_1} \ldots \partial_{\mu_{n}} %= q^{\mu_1} \ldots q^{\mu_{n}} \partial_{\mu_1} \ldots \partial_{\mu_{n}}
\ee
we have
\ba
D^{n+2}_w \left[ \Kwb  (q \cdot \partial)^{n} \right] & =&  D^{n+2}_w [ \Kwb q^{N}] \partial_{N} \\
& =& \sum_{m=0}^{n}\binom{n+2}{m}  [D^{n+2-m}_w \Kwb D^m_w q^{N}] \partial_{N}.
\ea
Note that the sum goes up to $m=n$ since $D^{n+1}_w q^{N}=0$ (as a consequence of $D^2_w q^\mu=0$). This implies that the derivatives acting on $\Kwb$ are at least of second order, which  allows  the use of identity (\ref{D2K}) leading to:
\be
D^{n-m+2}_w \Kwb = D^{n-m+1}_w \delta^{(2)}(w,z) \partial_E + D^{n-m}_w \delta^{(2)}(w,z)  E^{-1} \partial_z. \label{idK}
\ee 
Using these expressions, the integral in (\ref{TT}) localizes in $\ph= \qh$. The result can be written as
\be
\T= \T_1 + \T_2,
\ee
\ba
\T_1 & = & \sum_{m=0}^{n}\binom{n+2}{m} (-1)^{n-m+1}  D^{n-m+1}_z( T D^m_z q^{N}) \partial_E \partial_{N} \label{T1} \\
\T_2 & = & \sum_{m=0}^{n}\binom{n+2}{m} (-1)^{n-m}  D^{n-m}_z( T D^m_z q^{N}) E^{-1} \partial_z \partial_{N}, \label{T2}
\ea
corresponding to the first and second term in (\ref{idK}). In these and  following expressions it is  understood that $q^\mu$ is evaluated at $\qh=\ph$:
\be
q^\mu= (1, \ph).
\ee
We now simplify each term.  Applying general Leibinz rule to the $D^{n-m+1}_z$ derivative in (\ref{T1}) and using the identity 
\be
\sum_{m=0}^k (-1)^{n-m+1} \binom{n+2}{m} \binom{n-m+1}{n-k+1}  =(-1)^{n-k+1}
\ee
one arrives at:
\be
\T_1 = \sum_{k=0}^n (-1)^{n-k+1}D^{n-k+1}_z T D^k_z q^N \partial_E \partial_{N}.
\ee
Since $q^N$ is independent of $E$, the $\partial_E$ derivative can be permuted with $D^k_z q^N$. In subsection \ref{DkqNsec} we show that
\be
D^k_z q^N  \partial_{N} = \frac{n!}{(n-k)!} \partial^{n-k}_E E^{-k} D_z^k \label{DkqN}
\ee
and so we arrive at
\be
\T_1 =  (-1)^{n+1} \sum_{k=0}^n(-1)^k \frac{n!}{(n-k)!} D^{n-k+1}_z T \partial^{n-k+1}_E E^{-k} D_z^k. \label{T12}
\ee
Similar steps can  be carried out for $\T_2$. Expanding the derivatives in (\ref{T2}) and using the identity
\be
\sum_{m=0}^k (-1)^{n-m} \binom{n+2}{m} \binom{n-m}{n-k}  =(-1)^{n-k}(k+1)
\ee
one arrives at
\be
\T_2  =    \sum_{k=0}^n (-1)^{n-k}(k+1) D^{n-k}_z T D^k_z q^N  E^{-1} \partial_z \partial_{N} .\label{T22} 
\ee
In subsection \ref{DkqNsec} we show that
\be
D_z^k q^N  E^{-1} \partial_z \partial_{N} = \frac{n!}{(n-k)!} \partial^{n-k}_E E^{-(k+1)} D^{k+1}_z \label{DkqN2}
\ee
and so we obtain
\be
\T_2 = (-1)^{n} \sum_{k=0}^n (-1)^{k}\frac{n! (k+1) }{(n-k)!}D^{n-k}_z T \partial^{n-k}_E  E^{-k-1} D_z^{k+1}. \label{T23}
\ee
We finally combine both terms. Relabeling the summation index in (\ref{T23}) by $k \to k - 1$ and adding the result to (\ref{T12}) one arrives at
\be
 \T =  (-1)^{n+1} \sum_{k=0}^{n+1}(-1)^k \frac{(n+1)!}{(n+1-k)!} D^{n-k+1}_z T \partial^{n-k+1}_E E^{-k} D_z^k. \label{Tfinalapp}
\ee
One can verify that the operator satisfies
\be
\int d^2 \xh \int d E E \,  \psi(E,\xh) \T [ \phi(E,\xh) ]  = (-1)^{n+1} \int d^2 \xh \int d E E \, \T [\psi(E,\xh) ] \phi(E,\xh) ,
\ee
which was the required condition for the existence of a charge reproducing the factors in the soft theorems.

Doing the replacement $n \to n-1$ and including a multiplicative factor $ \frac{(-i)^{n+1}}{2}$ in (\ref{Tfinalapp}) one obtains (\ref{Tfinal}).
\subsubsection{Eqs. (\ref{DkqN}) and (\ref{DkqN2})} \label{DkqNsec}
Since $D^2_z q^\mu =0$, the expansion of $D^k_z q^N$ only includes terms where a derivative can at most hit once to each $q^{\mu_i}$. There is a total of  $\frac{n!}{(n-k)!}$ such terms, and they all give the same contribution in the contraction  $D^k_z q^N  \partial_{N}$  since the indices in  $\partial_{N}$ are totally symmetric. We can thus write
\be
D^k_z q^N  \partial_{N}  =\frac{n!}{(n-k)!} D_z q^{\mu_1} \ldots D_z q^{\mu_k} q^{\mu_{k+1}} \ldots q^{\mu_n} \partial_K \partial_{N-K}. \label{DkqN0}
\ee
We next show the identities:
\be
q^{\mu_1}\ldots q^{\mu_m} \partial_{\mu_1} \ldots \partial_{\mu_m} = \partial^m_E, \label{qmupartial}
\ee
\be
D_z q^{\mu_1}\ldots D_z q^{\mu_m} \partial_{\mu_1} \ldots \partial_{\mu_m} = E^{-m} D^m_z. \label{Dzqmupartial}
\ee
From which  (\ref{DkqN}) then follows.  Eqs. (\ref{qmupartial}), (\ref{Dzqmupartial}) can be shown by induction. Using (\ref{partialmu}) we have
\be
q^\mu \partial_\mu  =  \partial_0 + \partial_\rho = \partial_E 
\ee 
\be
D_z q^\mu \partial_\mu = E^{-1} D_z \label{Dzqmu}
\ee
where in (\ref{Dzqmu}) we used the on-shell condition $\rho=E$ and the identity $D_B \ph^i D^A \ph_i = \delta^A_B$. Assuming (\ref{qmupartial}) is true for $m-1$ we have:
\be
q^{\mu_1}\ldots q^{\mu_m} \partial_{\mu_1} \ldots \partial_{\mu_m} = q^{\mu_m} \partial^{m-1}_E \partial_{\mu_m} = \partial^{m-1}_E q^{\mu_m} \partial_{\mu_m}  =  \partial^m_E.
\ee
Similarly, assuming (\ref{Dzqmupartial}) is valid for $m-1$ one has:
\ba
D_z q^{\mu_1}\ldots D_z q^{\mu_m} \partial_{\mu_1} \ldots \partial_{\mu_m} & = & D_z q^{\mu_m} E^{-{(m-1)}} D^{m-1}_z \partial_{\mu_m} \\
& = &  E^{-{(m-1)}} D^{m-1}_z D_z q^{\mu_m} \partial_{\mu_m} \\
&=& E^{-m} D^m_z
\ea
where we could move $D_z q^{\mu_m}$ pass the derivatives since $D_z(D_z q^{\mu}) =0$.

To show Eq. (\ref{DkqN2}) we write $E^{-1} \partial_z = D_z q^{\mu_0} \partial_{\mu_0}$ so that: 
\ba
D^k_z q^N  E^{-1} \partial_z \partial_{N}  &= & \frac{n!}{(n-k)!}D_z q^{\mu_0}  \ldots D_z q^{\mu_k} q^{\mu_{k+1}} \ldots q^{\mu_n} \partial_{K+1} \partial_{N-K} \\
&= &  \frac{n!}{(n-k)!}\partial^{n-k}_E E^{-(k+1)} D^{k+1}_z.
\ea

\section{Appendix for section \ref{class-cons}}
\subsection{Identities on $\H$} \label{idsH}
\ba
\D_\alpha Y_\mu \D^\alpha Y^\nu & = & \delta^\nu_\mu - Y_\mu  Y^\nu \\
 \D_\alpha Y_\mu \D_\beta Y^\mu & = & h_{\alpha \beta}  \\
\D_\alpha \D_\beta Y_\mu & =&  - h_{\alpha \beta} Y_\mu .
\ea
These identities together with  $[\D^\beta,\D_\alpha] V_\beta  =  2 V_\alpha$ imply:
\be
\D^2 J^{\mu \nu}_{\alpha} = -2 J^{\mu \nu}_{\alpha}
\ee
\be
\D_\beta J^{\mu \nu}_{\alpha} \D^\beta (Y \cdot q) = - (Y \cdot q)J^{\mu \nu}_{\alpha} + (q^\mu \D_\alpha Y^\nu - (\mu \leftrightarrow \nu)).
\ee

\be
J^{\mu \nu}_A L_{\mu \nu}^B \to \delta^B_A \quad \text{for} \quad  \qh \to \xh,
\ee
\be
\D^2 f(Y \cdot q)  =  -(Y \cdot q)^2 f''(Y \cdot q)-3 (Y \cdot q)f'(Y \cdot q)  
\ee

\subsection{Eqs. (\ref{fallt1}) and (\ref{fallt2})} \label{fallt}
Let us introduce the notation
\be
\sigma: = Y \cdot q = -\t +\sqrt{1+\t^2} \xh \cdot \qh
\ee
Then the contraction of angular momenta in (\ref{candidate}) can be written as
\be
\frac{1}{2}J^{\mu \nu}_\alpha(y) L_{\mu \nu}^B(\qh) =  \sigma D_\alpha \bar{D}^B \sigma -  D_\alpha \sigma \bar{D}^{B}\sigma
\ee
where the barred derivative refers to the $\qh$ variable.  Using this expression in (\ref{candidate}), together with the identity $ \sigma \delta^{(m+1)}(\sigma) = -(m+1) \delta^{(m)}(\sigma)$ and some integration by parts one arrives at
\be
\overset{m}{F}_{\rho \alpha}(\t,\xh) =  - m \int d^2 \qh \delta^{(m)}(\sigma) D_\alpha \bar{D}^{B} \sigma  \overset{m}{V}_B+ D_\alpha \int d^2 \qh \delta^{(m-1)}(\sigma) \bar{D}^{B} \overset{m}{V}_B. \label{greenalpha}
\ee
We first focus on the $\alpha=\tau$ component. Introducing the notation
\ba
c & := & \xh \cdot \qh  \\ 
c_\t & := & \frac{\t}{\sqrt{1+\t^2}} \stackrel{\t \to \infty}{\approx} 1-\frac{1}{2\t^2}
\ea
so that
\be
\sigma= \sqrt{1+\t^2}(c-c_\t), \quad \partial_\t \sigma= c_\t c -1
\ee
one finds that for $\alpha=\t$ the two contributions of (\ref{greenalpha}) combine to give:
\be
\overset{m}{F}_{\rho \t}= \frac{1}{(1+\t^{2})^{m/2}} \partial_\t \int d^2 \qh \delta^{(m-1)}(c-c_\t) \bar{D}^B \overset{m}{V}_B \label{Fmtau}
\ee

\subsubsection{$\t \to \infty$ evaluation of integrals}
We need to evaluate integrals of the form
\be
I(\t) := \int d^2 \qh \delta^{(m)}(c-c_\t) f(\qh). \label{Itau}
\ee
If we parametrize the $\qh$ sphere by $(c,\phi)$ so that the sphere line element reads
 \be
 d s^2 = \frac{d c^2}{1-c^2}+(1-c^2) d \phi^2 \label{sphere}
 \ee
the integral becomes
\be
I(\t) = \int_{-1}^{1} dc \int_0^{2 \pi} d\phi  \delta^{(m)}(c-c_\t) f(c,\phi). 
\ee
To work out further the expression let us consider the case of an axially symmetric function $f(c,\phi)=f(c)$. Since we are  averaging over $\phi$, the result will also apply to the general (nonsymmetric) case.
For $f(c,\phi)=f(c)$ we  have:
\be
I(\t) = 2 \pi (-1)^{m} f^{(m)}(c_\t) \stackrel{\t \to \infty}{\approx}2 \pi (-1)^{m}\left( f^{(m)}(1) -\frac{1}{2 \t^2}f^{(m+1)}(1) \right) +O(\t^{-4}).
\ee
We finally need to express $f^{(m)}(1)$ in terms of powers of Laplacians. This can be done by writing the Laplacian in the coordinates (\ref{sphere}) at $c=1$:
\be
\Delta f(c)|_{c=1} = \partial_c (1-c^2) \partial_c f(c)|_{c=1}= - 2 f'(1).
\ee
By computing higher order Laplacians one can find expressions for the quantities $f^{(m)}(1)$ in terms of Laplacians. For instance by computing $\Delta^2 f:$
\be
\Delta^2 f(c)|_{c=1} = 4 f'(1)+8f''(1)
\ee
one concludes
\be
f''(1) = \frac{1}{8} (\Delta+2) \Delta f (c)|_{c=1}
\ee
Doing this to arbitrary order one finds\footnote{ Expression (\ref{fmDeltam}) was verified in Mathematica for given values of $m$ but we lack a rigorous proof of this relation.}
\be \label{fmDeltam}
f^{(m)}(1) = \Delta(m,1) f(c)|_{c=1}
\ee
where $\Delta(m,1) = \prod_{k=1}^m\Delta_k$ the differential operator introduced in section \ref{four}, Eq. (\ref{defDeltanm}). 
We thus obtain the following $\t \to \infty$ expansion for (\ref{Itau}):
\be
 \int d^2 \qh \delta^{(m)}(c-c_\t) f(\qh) \stackrel{\t \to  \infty}{\approx} 2 \pi (-1)^{m} \Delta(m,1)f(\xh) + \frac{\pi}{\t^2} (-1)^{m+1} \Delta(m+1,1)f(\xh) \label{formula}
\ee
Applying this formula to (\ref{Fmtau}) we obtain 
\be
\overset{m}{F}_{\rho \t}\stackrel{\t \to  \infty}{\approx} \frac{2 \pi (-1)^{m+1}}{\t^{m+3}} \Delta(m,1) D^B \overset{m}{V}_B. \label{Frhotauinf}
\ee
giving the desired fall-offs with leading term as in Eq. (\ref{fallt2}).

 $\overset{n}{F}_{\rho A}$ can be treated along similar lines. Here a difficulty stems from the need to work with tensorial indices. A way  to deal with this problem is to decompose the vector field $\overset{m}{V}_A$ into grad plus curl pieces, $\overset{m}{V}_A= \partial_A f + \e_A^B \partial_B g$ and find scalar expressions involving $f$ and $g$. This allows one to use the formula (\ref{formula}) (valid for scalars) to arrive at the desired asymptotic form as given in Eq. (\ref{fallt1}).

\end{document}